\documentclass[english,onecolumn]{IEEEtran}
\usepackage[T1]{fontenc}
\usepackage[latin9]{inputenc}
\usepackage{babel}
\usepackage{amsmath}
\usepackage{amsthm}
\usepackage{amssymb}
\usepackage{graphicx}
\usepackage{tablefootnote}
\usepackage{esint}
\usepackage[unicode=true]
 {hyperref}

\makeatletter

\providecommand{\tabularnewline}{\\}

\theoremstyle{plain}
\newtheorem{thm}{\protect\theoremname}
\theoremstyle{plain}
\newtheorem{cor}[thm]{\protect\corollaryname}

\usepackage{mathrsfs}
\usepackage{algorithm,algpseudocode}

\usepackage{colortbl}
\definecolor{lightgray}{rgb}{0.9,0.9,0.9}
\definecolor{lightred}{rgb}{1,0.8,0.8}
\definecolor{lightgreen}{rgb}{0.6,1,0.6}
\definecolor{lightyellow}{rgb}{1,1,0.5}
\definecolor{lightgrey}{rgb}{0.8,0.8,0.8}

\allowdisplaybreaks[1]

\makeatother

\providecommand{\corollaryname}{Corollary}
\providecommand{\theoremname}{Theorem}

\begin{document}
\title{The Undecidability of Conditional Affine Information Inequalities
and Conditional Independence Implication with a Binary Constraint}
\author{Cheuk Ting Li, \textit{Member, IEEE}\thanks{This paper was presented
in part at the 2021 IEEE Information Theory Workshop. \newline C.
T. Li is with the Department of Information Engineering, The Chinese
University of Hong Kong (e-mail: ctli@ie.cuhk.edu.hk).}\\
}
\maketitle
\begin{abstract}
We establish the undecidability of conditional affine information
inequalities, the undecidability of the conditional independence implication
problem with a constraint that one random variable is binary, and
the undecidability of the problem of deciding whether the intersection
of the entropic region and a given affine subspace is empty. This
is a step towards the conjecture on the undecidability of conditional
independence implication. The undecidability is proved via a reduction
from the periodic tiling problem (a variant of the domino problem).
Hence, one can construct examples of the aforementioned problems that
are independent of ZFC (assuming ZFC is consistent). 
\end{abstract}


\medskip{}

\section{Introduction}

The problem of characterizing the entropic region $\Gamma_{n}^{*}$
and the almost-entropic region $\overline{\Gamma_{n}^{*}}$ (the closure
of $\Gamma_{n}^{*}$) is a fundamental problem in information theory
\cite{zhang1997non,yeung1997framework,zhang1998characterization}.
Its applications include network coding \cite{yeung2008information,chan2008dualities,yan2012implicit,apte2014algorithms,apte2016explicit},
secret sharing \cite{csirmaz1997size,metcalf2011improved,kaced2012secret,martin2015secret,farras2018improving,gurpinar2019use},
group theory \cite{chan2002relation}, and automated proofs of capacity
regions in multiuser coding settings \cite{yeung1996itip,gattegno2016fourier,li2021automated}.

Pippenger \cite{pippenger1986laws} raised the question whether all
inequalities among entropy and mutual information of random variables
can be deduced from $I(X;Y|Z)\ge0$ (i.e., Shannon-type inequalities).
This was answered by Zhang and Yeung, who showed the first conditional
non-Shannon-type inequality (i.e., showing that an inequality on entropy
terms is implied by a collection of other such inequalities, but this
implication cannot be proved using Shannon-type inequalities) in \cite{zhang1997non},
and the first unconditional non-Shannon-type inequality in \cite{zhang1998characterization}.
More non-Shannon-type inequalities were discovered later in \cite{makarychev2002new,dougherty2006six,matus2007infinitely,xu2008projection,dougherty2011non}.
In particular, Mat{\'u}{\v{s}} \cite{matus2007infinitely} showed
that $\overline{\Gamma_{n}^{*}}$ is not polyhedral. Chan and Grant
\cite{chan2008non} proved a non-linear information inequality, and
conjectured that $\overline{\Gamma_{n}^{*}}$ is semialgebraic. It
was shown by G{\'o}mez, Mej{\'\i}a and Montoya \cite{gomez2014network}
that $\Gamma_{n}^{*}$ for $n\ge3$ is not semialgebraic, though it
is still unknown whether $\overline{\Gamma_{n}^{*}}$ is semialgebraic
in general. 

To study unconditional linear information inequalities, it suffices
to consider $\overline{\Gamma_{n}^{*}}$, which is a convex cone.
Nevertheless, if conditional information inequalities are of interest
(e.g. for the conditional independence implication problem and secret
sharing \cite{kaced2012secret}), we have to work with $\Gamma_{n}^{*}$
which is non-convex. Kaced and Romashchenko \cite{kaced2013conditional}
showed the existence of conditional information inequalities that
are not implied by unconditional linear information inequalities,
and conditional information inequalities that are valid in $\Gamma_{n}^{*}$
but not in $\overline{\Gamma_{n}^{*}}$ (also see \cite{kaced2012non}).
This suggests that characterizing conditional information inequalities
might be harder than characterizing unconditional ones (which is already
extremely hard). 

While there are algorithms that attempt to verify conditional information
inequalities (e.g. \cite{yeung1996itip,pulikkoonattuxitip,ho2020proving}),
most of them only take Shannon-type inequalities into account, and
hence fail to verify true non-Shannon-type inequalities. There are
algorithms capable of verifying some non-Shannon-type inequalities
\cite{xu2008projection,dougherty2011non,farras2018improving,gurpinar2019use,li2021automated},
though there is no known algorithm that is capable of verifying \emph{every}
true conditional information inequality. It was unknown whether such
algorithm can exist, that is, whether the problem of conditional information
inequality is algorithmically decidable. See \cite{dougherty2009undecidable,gomez2014network,gomez2017defining,gomez2018theory,kuhne2019representability,khamis2020decision}
for partial results on the decidability/undecidability of information
inequalities and network coding. 

A closely related problem is the conditional independence implication
problem \cite{dawid1979conditional,spohn1980stochastic,mouchart1984note,pearl1987graphoids},
which is to decide whether a statement on the conditional independence
among several random variables follows from a list of other such statements.
Since conditional independence can be expressed as $I(X;Y|Z)=0$,
this problem can be reduced to conditional information inequalities.
Pearl and Paz \cite{pearl1987graphoids} introduced a set of axioms
(the semi-graphoid axioms) capable of solving a subset of conditional
independence implication problems. This set of axioms was shown to
be incomplete by Studen\'y \cite{studeny1989multiinformation}, who
later showed in \cite{studeny1990conditional} that there is no finite
axiomization of probabilistic conditional independence in general. 

Nevertheless, conditional independence implication among random variables
with fixed cardinalities can be decided algorithmically, as shown
by Niepert \cite{niepert2012logical}. Hannula et al. \cite{hannula2019facets}
showed that if all random variables are binary, then the problem is
in EXPSPACE. Khamis, Kolaitis, Ngo and Suciu \cite{khamis2020decision}
proved that the general conditional independence implication problem
is at most in $\Pi_{1}^{0}$ in the arithmetical hierarchy. It is
unknown whether the conditional independence implication problem is
decidable in general. See \cite{geiger1991axioms,geiger1993logical,bouckaert2007racing,niepert2010logical,bouckaert2010efficient,de2012stable,niepert2013conditional,gyssens2014completeness}
for other partial results on the decidability/undecidability of conditional
independence implication. 

In this paper, which is an extended version of \cite{li2021undecidabilityitw}\footnote{Note that \cite{li2021undecidabilityitw} only contains a brief version
of the proof of the undecidability of the affine subspace intersection
problem, and does not contain the proof for the conditional independence
implication result.}, we give a partial negative answer to the question on the algorithmic
decidability of conditional information inequalities. We show that
the problem of deciding whether the intersection of the entropic region
and a given affine subspace is empty is undecidable. More precisely,
we show the undecidability of the problem of deciding whether there
exists $\mathbf{v}\in\Gamma_{n}^{*}$ such that $\mathbf{A}\mathbf{v}=\mathbf{b}$,
where $\mathbf{A}\in\mathbb{Q}^{m\times(2^{n}-1)}$, $\mathbf{b}\in\mathbb{Q}^{m}$
are given rational matrix and vector, and $n,m\ge0$ are given integers.
We remark that affine or nonhomogeneous inequalities have appeared
in existential results about random variables before (e.g. \cite{li2021infiniteisit,sfrl_trans}),
so the study of affine constraints is not unnatural.

We also prove the undecidability of the conditional independence implication
problem with a binary constraint. Consider random variables $X_{1},\ldots,X_{n}$,
$n\ge2$. This problem is undecidable: deciding whether $I(X_{A_{j}};X_{B_{j}}|X_{C_{j}})=0$
for $j=1,\ldots,m$ and $X_{1}$ has cardinality at most $2$ (i.e.,
$X_{1}$ is binary) implies that $I(X_{1};X_{2})=0$, where $A_{j},B_{j},C_{j}\subseteq\{1,\ldots,n\}$
are three given disjoint sets\footnote{Some authors have considered the non-disjoint case \cite{dawid1979conditional,van1998informational},
whereas some authors require those sets to be disjoint \cite{pearl1987graphoids,studeny1997semigraphoids}.
We prove in Section \ref{subsec:disjoint_case} that the two cases
are actually Turing equivalent, and non-disjoint conditional independence
relations can be expressed as disjoint conditional independence relations.} for each $j=1,\ldots,m$, and $X_{A}:=\{X_{i}\}_{i\in A}$ (see Corollary
\ref{cor:imp_disjoint}). To the best of the author's knowledge, this
is the first undecidability result about probabilistic conditional
independence\footnote{We remark that the embedded multivalued dependencies implication problem,
a different but related problem, was shown to be undecidable in \cite{herrmann1995undecidability,herrmann2006corrigendum}.
We also remark that the capacity of finite state machine channel is
uncomputable \cite{elkouss2018memory} (also see \cite{agarwal2018non,boche2020shannon}),
though this is unrelated to the setting in this paper.}, and hence is a step towards the conjecture on the undecidability
of conditional independence implication. This undecidability result
is perhaps surprising, considering that the conditional independence
implication problem when the cardinalities of all random variables
are bounded is decidable \cite{niepert2012logical} (also see \cite{hannula2019facets}),
so it is counterintuitive that bounding the cardinality of only one
random variable makes the problem undecidable (see Table \ref{tab:ci_decidability}).

\smallskip{}

Furthermore, we prove that the following problems are undecidable:

\smallskip{}

\begin{itemize}
\item (One linear equality and one affine equality) Deciding whether there
exists $\mathbf{v}\in\Gamma_{n}^{*}$ such that $\mathbf{a}^{T}\mathbf{v}=0$
and $v_{\{1\}}=1$, where $\mathbf{a}\in\mathbb{Q}^{2^{n}-1}$ and
$v_{\{1\}}$ denotes the entry of $\mathbf{v}$ that represents the
entropy of the first random variable.
\end{itemize}
\medskip{}

\begin{itemize}
\item (Conditional affine information inequality) Deciding the truth value
of the conditional affine information inequality in the form 
\[
\mathbf{v}\in\Gamma_{n}^{*}\;\wedge\;\mathbf{a}^{T}\mathbf{v}\le0\;\wedge\;v_{\{1\}}\le1\;\Rightarrow\;v_{\{1\}}=0,
\]
where $\mathbf{a}\in\mathbb{Q}^{2^{n}-1}$ (i.e., deciding whether
$\mathbf{v}\in\Gamma_{n}^{*}$, $\mathbf{a}^{T}\mathbf{v}\le0$ and
$v_{\{1\}}\le1$ implies $v_{\{1\}}=0$).
\end{itemize}
\medskip{}

\begin{itemize}
\item (Boolean information constraint) Deciding the truth value of the Boolean
information constraint \cite{khamis2020decision} (with strict affine
inequality constraints) in the form $\forall\mathbf{v}\in\Gamma_{n}^{*}:\,\mathbf{a}_{1}^{T}\mathbf{v}>b_{1}\;\vee\,\cdots\,\vee\;\mathbf{a}_{N}^{T}\mathbf{v}>b_{N}$. 
\end{itemize}
\medskip{}
We will show the undecidability of entropic region via a reduction
from the periodic tiling problem. The domino problem introduced by
Wang \cite{wang1961proving} concerns the problem of tiling the plane
with unit square tiles. Given a set of tiles, where each tile has
four colored edges, the problem is to decide whether it is possible
to tile the plane using this set of tiles, such that touching edges
of adjacent tiles have the same color. Each tile in the set can be
used an unlimited number of times, though no rotation or reflection
is allowed. It was proved by Berger \cite{berger1966undecidability}
that deciding whether a finite set of tiles can tile the plane is
undecidable by simulating a Turing machine using the tiles, and hence
undecidability follows from the undecidability of the halting problem
\cite{turing1937computable}. Moreover, Gurevich and Koryakov \cite{gurevich1972remarks}
showed that the problem of deciding whether a finite set of tiles
can tile the plane periodically (i.e., there exists a positive integer
$N$ such that the tiling remains the same after shifting upward by
$N$ or leftward by $N$, or equivalently, it can tile a torus) is
also undecidable. This is also proved by Mazoyer and Rapaport \cite{mazoyer1999global}
for NW-deterministic tiles\footnote{In \cite{mazoyer1999global}, for any Turing machine, a set of NW-deterministic
tiles is constructed such that the set of tiles admits a periodic
tiling if and only if the Turing machine halts.}. 

As a consequence, one can explicitly construct $\mathbf{A}\in\mathbb{Q}^{m\times(2^{n}-1)}$,
$\mathbf{b}\in\mathbb{Q}^{m}$ such that the non-existence of $\mathbf{v}\in\Gamma_{n}^{*}$
with $\mathbf{A}\mathbf{v}=\mathbf{b}$ is unprovable in ZFC (assuming
ZFC is consistent). Also, one can construct an example of a conditional
independence implication problem with a binary constraint which is
unprovable in ZFC. This is due to the fact that one can construct
a Turing machine such that whether it halts is independent of ZFC,
assuming ZFC is consistent (see e.g. \cite{michel2009busy,yedidia2016relatively}),
and that the set of tiles in \cite{gurevich1972remarks,mazoyer1999global}
and the $\mathbf{A},\mathbf{b}$ in this paper (Theorem \ref{thm:undecidable})
are given constructively. Finding the smallest $\mathbf{A}\in\mathbb{Q}^{m\times(2^{n}-1)}$,
$\mathbf{b}\in\mathbb{Q}^{m}$ such that $\exists\mathbf{v}\in\Gamma_{n}^{*}:\mathbf{A}\mathbf{v}=\mathbf{b}$
is independent of ZFC is left for future studies.

We remark that the decidability of the following problems remain open
(also see Table \ref{tab:ci_decidability}):
\begin{itemize}
\item Deciding the truth value of a conditional linear information inequality
(our construction requires one affine inequality condition).
\item Deciding the truth value of a conditional affine information inequality
with only one equality/inequality condition, e.g., $\mathbf{v}\in\Gamma_{n}^{*}\,\wedge\,\mathbf{a}^{T}\mathbf{v}\le b\,\Rightarrow\,\mathbf{c}^{T}\mathbf{v}\le d$
(our construction requires one linear equality and one affine equality
for the condition).
\item Deciding the truth value of a conditional affine information inequality
for a fixed $n\ge4$.
\item Deciding whether an affine subspace intersects the almost-entropic
region $\overline{\Gamma_{n}^{*}}$ (our arguments rely on the discreteness
of the boundary of $\Gamma_{n}^{*}$).
\item The conditional independence implication problem without any cardinality
constraint.
\end{itemize}
\medskip{}

\begin{table}
\begin{centering}
\begin{tabular}{|c|c|c|c|}
\hline 
 & $\begin{array}{c}
\text{No cardinality}\\
\text{bounds}
\end{array}$  & $\begin{array}{c}
\text{Cardinality bounds}\\
\text{on some variables}
\end{array}$  & $\begin{array}{c}
\text{Cardinality bounds}\\
\text{on all variables}
\end{array}$\tabularnewline
\hline 
\hline 
$\begin{array}{c}
\text{Disjoint}\;A_{j},B_{j},\\
C_{j}=\emptyset
\end{array}$ & \cellcolor{lightgreen}Decidable \cite{geiger1991axioms,matuvs1994stochastic} & \cellcolor{lightgreen}Decidable \cite{geiger1991axioms}\tablefootnote{This follows from \cite[Theorem 3]{geiger1991axioms} which uses only
binary random variables to show completeness, and that random variables
with cardinality bound 1 can be ignored.} & \cellcolor{lightgreen}Decidable \cite{geiger1991axioms,niepert2012logical}\tabularnewline
\hline 
$\begin{array}{c}
\text{Disjoint}\;A_{j},B_{j},C_{j},\\
A_{j}\cup B_{j}\cup C_{j}=[n]\\
\text{(saturated CI)}
\end{array}$ & \cellcolor{lightgreen}Decidable \cite{malvestuto1992unique,geiger1993logical} & \cellcolor{lightgreen}Decidable \cite{geiger1993logical}\tablefootnote{This follows from \cite[Theorem 15]{geiger1993logical} which uses
only binary random variables to show the completeness of the semi-graphoid
axioms, and that random variables with cardinality bound 1 can be
ignored.} & \cellcolor{lightgreen}Decidable \cite{geiger1993logical,niepert2012logical}\tabularnewline
\hline 
$\begin{array}{c}
\text{General disjoint}\\
A_{j},B_{j},C_{j}
\end{array}$ & Unknown & \cellcolor{lightred}$\begin{array}{c}
\text{Undecidable}\\
\text{(this paper)}
\end{array}$  & \cellcolor{lightgreen}Decidable \cite{niepert2012logical}\tabularnewline
\hline 
\end{tabular}
\par\end{centering}
\medskip{}

\caption{\label{tab:ci_decidability}The decidability status of the conditional
independence implication problem under various conditions on the set
of random variables $A_{j},B_{j},C_{j}$ and cardinality bounds. \textquotedblleft Cardinality
bounds on some variables\textquotedblright{} means that we restrict
$\mathrm{card}(X_{i})\le k_{i}$ for given $k_{1},\ldots,k_{n}\in\{1,2,\ldots\}\cup\{\infty\}$.
\textquotedblleft Cardinality bounds on all variables\textquotedblright{}
means $k_{i}$ is finite for all $i$.}
\end{table}

This paper is organized as follows. In Section \ref{sec:lp}, we review
the entropic region. In Section \ref{sec:aeip}, we introduce the
concept of affine existential information predicate (an extension
of existential information predicate \cite{li2021automated}), the
main tool in our proof. In Section \ref{sec:tile}, we prove the main
result on the undecidability of the intersection of $\Gamma_{n}^{*}$
and an affine subspace. In Section \ref{sec:ci}, we prove the undecidability
of the satisfiability and implication problem of conditional independence
with a binary constraint. In Section \ref{sec:related}, we show the
undecidability of related problems as corollaries of our main result.

\medskip{}

\subsection*{Notations}

Throughout this paper, entropy is in bits, and $\log$ is to the base
$2$. The binary entropy function (i.e., the entropy of $\mathrm{Bern}(t)$)
is denoted as $H_{b}(t)$. Given a discrete random variable $X\in\mathcal{X}$,
its cardinality (the size of its support) is denoted as $\mathrm{card}(X):=|\{x\in\mathcal{X}:\,\mathbf{P}(X=x)>0\}|$.
The set of rational real numbers is denoted as $\mathbb{Q}$. Given
propositions $P,Q$, the logical conjunction (i.e., AND) is denoted
as $P\wedge Q$. 

Given vectors $\mathbf{v},\mathbf{w}\in\mathbb{R}^{k}$, $\mathbf{v}\succeq\mathbf{w}$
means $v_{i}\ge w_{i}$ for any $i$. The sign function is denoted
as $\mathrm{sgn}(t)=\mathbf{1}\{t>0\}-\mathbf{1}\{t<0\}$. We write
$[a..b]:=\mathbb{Z}\cap[a,b]$, $[n]:=\{1,\ldots,n\}$. We write $X_{a}^{b}:=(X_{a},X_{a+1},\ldots,X_{b})$,
$X^{n}:=X_{1}^{n}$. For finite set $S\subseteq\mathbb{N}$, write
$X_{S}:=(X_{a_{1}},\ldots,X_{a_{k}})$, where $a_{1},\ldots,a_{k}$
are the elements of $S$ in ascending order. We usually use $\mathbf{A}$
for matrix, $\mathbf{a},\mathbf{b},\mathbf{c},\mathbf{d}$ for column
vectors, $X,Y,Z,U,V,W$ for random variables, and $S$ for sets.

\medskip{}

\section{Entropic Region\label{sec:lp}}

In this section, we briefly review the definition of the entropic
region \cite{zhang1997non,yeung1997framework,zhang1998characterization}.
For a sequence of discrete random variables with finite entropies
$X^{n}=(X_{1},\ldots,X_{n})$, its entropic vector \cite{zhang1997non}
is defined as $\mathbf{h}(X^{n})=\mathbf{h}\in\mathbb{R}^{2^{n}-1}$,
where the entries of $\mathbf{h}$ are indexed by nonempty subsets
of $[n]$, and $\mathbf{h}_{S}:=H(X_{S})$ (where $S\subseteq[n]$)
is the joint entropy of $\{X_{i}\}_{i\in S}$. The entropic region
\cite{zhang1997non} is the region of entropic vectors
\[
\Gamma_{n}^{*}:=\bigcup_{p_{X^{n}}}\{\mathbf{h}(X^{n})\}
\]
over all discrete joint distributions $p_{X^{n}}$. The problem of
characterizing $\Gamma_{n}^{*}$ for $n\ge4$ is open. A polyhedral
outer bound $\Gamma_{n}$ characterized by Shannon-type inequalities
is given in \cite{zhang1998characterization}, which is the basis
of the linear program for verifying linear information inequalities
in \cite{yeung1996itip,yeung1997framework}.

\medskip{}

\section{Affine Existential Information Predicate\label{sec:aeip}}

The concept of existential information predicate (EIP) was studied
systematically in \cite{li2021automated}. Before that, several existential
statements on random variables and entropy terms have been studied,
for example, the copy lemma \cite{zhang1998characterization,dougherty2011non},
strong functional representation lemma \cite{harsha2010communication,sfrl_trans},
double Markov property \cite{csiszar2011information}, and the infinite
divisibility of information \cite{li2020infinite}.

Here we introduce an extension of EIP to allow affine inequalities,
called \emph{affine existential information predicate} (AEIP). An
AEIP is a predicate on the random sequence $X^{n}$ in the form
\[
\exists U^{l}\!:\mathbf{A}\mathbf{h}(X^{n},U^{l})\succeq\mathbf{b},
\]
where $\mathbf{A}\in\mathbb{R}^{m\times(2^{n+l}-1)}$, $\mathbf{b}\in\mathbb{R}^{m}$,
$m\ge0$. The ``$\exists U^{l}$'' should be interpreted as ``$\exists p_{U^{l}|X^{n}}$'',
i.e., there exists random sequence $U^{l}$ dependent on $X^{n}$
such that $\mathbf{A}\mathbf{h}(X^{n},U^{l})\succeq\mathbf{b}$. Denote
the above predicate as $\mathrm{AEIP}_{n,\mathbf{A},\mathbf{b}}(X^{n})$
(note that the value of $l$ can be deduced from $n$ and the width
of $\mathbf{A}$). 

While affine inequalities among entropy might seem unnatural since
they are nonhomogeneous (considering most previous works on information
inequalities are about linear inequalities, and the nonlinear inequality
in \cite{chan2008non} is homogeneous), it is not uncommon to have
inequalities involving both entropy and constant terms in information
theory, e.g. the expected length of Huffman code, and the Knuth-Yao
scheme for random number generation \cite{knuth1976complexity}. Some
specific examples of AEIPs have been studied previously. For example,
the approximate infinite divisibility of information \cite{li2021infiniteisit}
states that the following AEIP holds for any random variable $X$
and any $l\ge1$ (slightly relaxed since \cite{li2021infiniteisit}
requires $U_{1},\ldots,U_{l}$ to be i.i.d.):
\begin{align*}
\exists U^{l}: & \,U_{1}\perp\!\!\!\perp\cdots\perp\!\!\!\perp U_{l}\,\wedge\,H(X|U^{l})=0\\
 & \wedge\,H(U_{1})=\cdots=H(U_{l})\le\frac{1.59}{l}H(X)+2.43.
\end{align*}
Another example is the strong functional representation lemma \cite{sfrl_trans,li2021unified},
which states that the following holds for any random variables $X,Y$:
\begin{align*}
\exists U: & \,U\perp\!\!\!\perp X\,\wedge\,H(Y|X,U)=0\\
 & \wedge\,H(Y|U)\le I(X;Y)+\log(I(X;Y)+1)+4.
\end{align*}
While this is not itself an AEIP, it implies the following AEIP holds
for any fixed $\gamma\ge0$ and random variables $X,Y$:
\begin{align*}
\exists U: & \,U\perp\!\!\!\perp X\,\wedge\,H(Y|X,U)=0\\
 & \wedge\,H(Y|U)\le I(X;Y)+\frac{\log e}{\gamma+1}\left(I(X;Y)-\gamma\right)+\log(\gamma+1)+4.
\end{align*}
Also see \cite{makkuva2016additive} for some affine inequalities
on the entropy of sums of random variables. These suggest that homogeneity
is not a property we should expect from inequalities on entropy in
general, and affine inequalities are not unnatural. We remark that
AEIP can be handled by the Python Symbolic Information Theoretic Inequality
Prover in \cite{li2021automated} \footnote{Source code of Python Symbolic Information Theoretic Inequality Prover
(PSITIP) is available at \href{https://github.com/cheuktingli/psitip}{https://github.com/cheuktingli/psitip}}. 

Our goal is to show that the problem of deciding the truth value of
$\mathrm{AEIP}_{0,\mathbf{A},\mathbf{b}}(\emptyset)$ for a given
rational matrix $\mathbf{A}$ and rational vector $\mathbf{b}$ is
undecidable, where $\emptyset$ means $n=0$ ($X^{n}$ is empty).

We study some composition rules of AEIPs. For two AEIPs $\mathrm{AEIP}_{n,\mathbf{A},\mathbf{b}}(X^{n})$,
$\mathrm{AEIP}_{n,\mathbf{C},\mathbf{d}}(X^{n})$, their conjunction
(i.e., ``and'') is
\begin{align*}
\exists U^{l},V^{l'}\!: & \mathbf{A}\mathbf{h}(X^{n},U^{l})\succeq\mathbf{b}\\
 & \wedge\,\mathbf{C}\mathbf{h}(X^{n},V^{l'})\succeq\mathbf{d},
\end{align*}
which is clearly still an AEIP. Also, given an AEIP $\mathrm{AEIP}_{n+l,\mathbf{A},\mathbf{b}}(X^{n+l})$,
a predicate on $X^{n}$ in the form:
\begin{align*}
\exists U^{l}\!: & \mathrm{AEIP}_{n+l,\mathbf{A},\mathbf{b}}(X^{n},U^{l})\\
 & \wedge\,\mathbf{C}\mathbf{h}(X^{n},U^{l})\succeq\mathbf{d}
\end{align*}
is also an AEIP. 

We will use these composition rules to construct a class of undecidable
AEIPs in the next section. Note that while AEIP is quite general,
we will only use AEIPs defined using conditional independence constraints
(i.e., $I(X;Y|Z)=0$) and cardinality constraints \eqref{eq:unif_force}
throughout the proof. This will be important for results on conditional
independence implication in Section \ref{sec:ci}.

\medskip{}

\section{Construction of Tiling\label{sec:tile}}

In this section, we establish our main result, which is the undecidability
of AEIPs with rational coefficients.
\begin{thm}
\label{thm:undecidable}The problem of deciding the truth value of
$\mathrm{AEIP}_{\mathbf{A},\mathbf{b}}(\emptyset)$ (i.e., $\exists U^{l}\!:\mathbf{A}\mathbf{h}(U^{l})\succeq\mathbf{b}$)
given $\mathbf{A}\in\mathbb{Q}^{m\times(2^{l}-1)}$, $\mathbf{b}\in\mathbb{Q}^{m}$,
$l,m\ge0$ is algorithmically undecidable. Therefore, the problem
of deciding the truth value of $\exists U^{l}\!:\mathbf{A}\mathbf{h}(U^{l})=\mathbf{b}$
is also undecidable.
\end{thm}
The proof of Theorem \ref{thm:undecidable} is divided into several
steps presented in the following subsections.\medskip{}

\subsection{Construction of Tori/Grid}

It was shown in \cite{zhang1997non} that if $H(Y_{1}|Y_{2},Y_{3})=H(Y_{2}|Y_{1},Y_{3})=H(Y_{3}|Y_{1},Y_{2})=I(Y_{1};Y_{2})=I(Y_{1};Y_{3})=I(Y_{2};Y_{3})=0$,
then $Y_{1}$ is uniformly distributed over its support (also true
for $Y_{2},Y_{3}$), and they have the same cardinality. Write this
as the AEIP:
\begin{align}
 & \!\!\!\!\mathrm{TRIPLE}(Y_{1},Y_{2},Y_{3}):\nonumber \\
 & H(Y_{1}|Y_{2},Y_{3})=H(Y_{2}|Y_{1},Y_{3})=H(Y_{3}|Y_{1},Y_{2})\nonumber \\
 & =I(Y_{1};Y_{2})=I(Y_{1};Y_{3})=I(Y_{2};Y_{3})=0.\label{eq:triple}
\end{align}
Note that we use the notation ``$\mathrm{P}(X):\text{(formula with \ensuremath{X} as free variable)}$''
to define a predicate $\mathrm{P}$ instead of writing ``$\mathrm{P}(X)=\text{(formula)}$''
to avoid having to write extra parentheses (e.g. we can write $\mathrm{P}(X):H(X)=0$
instead of $\mathrm{P}(X)=(H(X)=0)$). Therefore, the predicate that
$X$ is uniformly distributed over its support can be expressed as
the following AEIP:
\begin{align*}
\mathrm{UNIF}(X):\,\exists U_{1},U_{2}:\, & \mathrm{TRIPLE}(X,U_{1},U_{2}).
\end{align*}
The predicate that $X$ is uniform with cardinality $k\ge2$ is given
by the AEIP:
\begin{equation}
\mathrm{UNIF}_{k}(X):\,\mathrm{UNIF}(X)\,\wedge\,\alpha_{k}\le H(X)\le\alpha_{k+1},\label{eq:unif_force}
\end{equation}
where $\alpha_{k}$ is a rational number such that $\log(k-1)<\alpha_{k}<\log k$. 

Given this, one natural approach to prove the desired undecidability
result is to show a reduction from the problem of deciding whether
a Diophantine equation (a polynomial equation with integer coefficients
and variables) has a solution, which is undecidable by Matiyasevich's
theorem \cite{matijasevic1970enumerable}. We can represent the positive
integer $k$ by a uniform random variable with cardinality $k$. It
is possible to define equality, multiplication and comparison over
positive integers using AEIPs (see Section \ref{subsec:nondisjoint}).
Nevertheless, it is uncertain whether addition (or the successor function)
over positive integers can be defined using AEIP (note that addition
of entropy corresponds to multiplication of cardinality), and hence
Matiyasevich's theorem cannot be applied. We will instead show a reduction
from the periodic tiling problem \cite{gurevich1972remarks}, another
undecidable (and seemingly less related) problem.

Given two random variables $X_{1},X_{2}$, the predicate that they
are uniform with the same cardinality, the pair $(X_{1},X_{2})$
is uniformly distributed over its support, and all vertices in their
characteristic bipartite graph (i.e., a graph with edge $(x_{1},x_{2})$
if and only if $p_{X_{1},X_{2}}(x_{1},x_{2})>0$) have degree $2$
(i.e., the bipartite graph consists of disjoint cycles) can be given
by the AEIP:
\begin{align*}
\mathrm{CYCS}(X_{1},X_{2}):\, & \exists U:\,\mathrm{UNIF}(X_{1})\,\wedge\,\mathrm{UNIF}(X_{2})\,\wedge\,\mathrm{UNIF}_{2}(U)\\
 & \wedge\,I(X_{1};U)=I(X_{2};U)=0\\
 & \wedge\,H(X_{1}|X_{2},U)=H(X_{2}|X_{1},U)=0\\
 & \wedge\,H(U|X_{1},X_{2})=0.
\end{align*}
This can be proved by observing $I(X_{1};U)=H(X_{2}|X_{1},U)=0$ implies
that $p_{X_{2}|X_{1}=x_{1}}$ is degenerate or uniform over two values
for any $x_{1}$, and the degenerate case is impossible since $H(U|X_{1},X_{2})=0$.
Hence all vertices have degree $2$. For the other direction, if the
bipartite graph consists of disjoint cycles, then we can color the
edges in two colors so that no two edges sharing a vertex have the
same color. The value of $U$ corresponds to the color of the edge.

\begin{figure}
\begin{centering}
\includegraphics[scale=0.8]{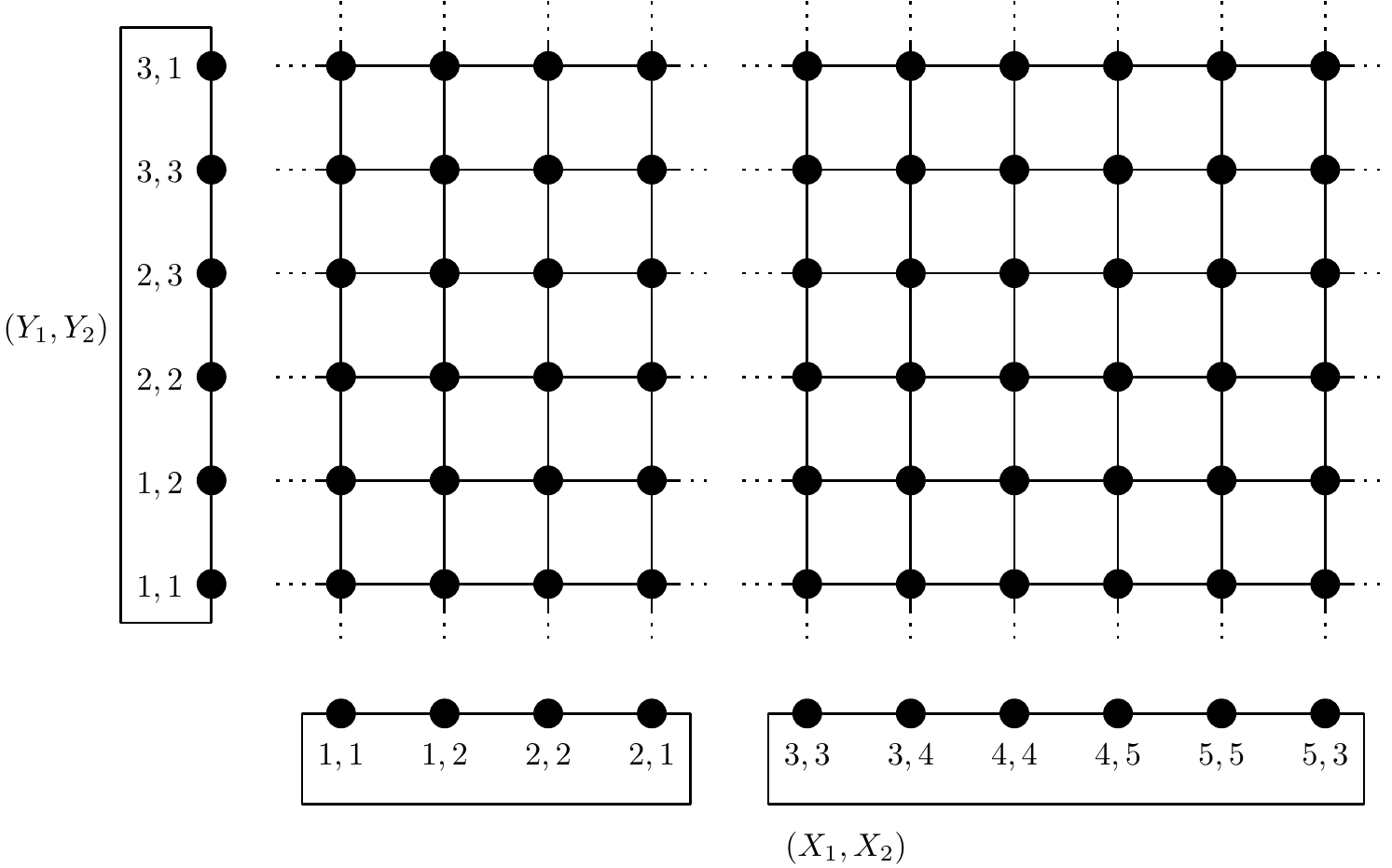}
\par\end{centering}
\caption{\label{fig:tori1}The tori constructed. The labels on the x-axis give
the support of $(X_{1},X_{2})$ (consisting of two cycles), whereas
the labels on the y-axis give the support of $(Y_{1},Y_{2})$ (consisting
of one cycle). Their product is a collection of two tori. }

\end{figure}

Tori can be formed by taking two independent copies of cycles. Define
the AEIP
\begin{align*}
\mathrm{TORI}(X^{2},Y^{2}):\, & \mathrm{CYCS}(X^{2})\,\wedge\,\mathrm{CYCS}(Y^{2})\\
 & \wedge\,I(X^{2};Y^{2})=0.
\end{align*}
Note that the characteristic bipartite graph between $(X_{1},Y_{1})$
and $(X_{2},Y_{2})$ is the disjoint union of a collection of tori
(product of cycles). Each vertex of the tori is represented as a quadruple
$(x_{1},x_{2},y_{1},y_{2})$ in the support of $(X_{1},X_{2},Y_{1},Y_{2})$.
Refer to Figure \ref{fig:tori1}. We will use the tori as the grid
for the tiling problem.

\medskip{}

\subsection{Construction of Colors}

We now describe a way to assign colors to the vertices of the tori.
Intuitively, the color of a vertex can be represented as a random
variable (which has values corresponding to colors). Nevertheless,
this does not work since the values of a random variable cannot be
identified using entropy, which is invariant under relabeling. Hence,
we have to introduce a way to encode color information using conditional
independence.

\begin{figure}
\begin{centering}
\includegraphics[scale=0.9]{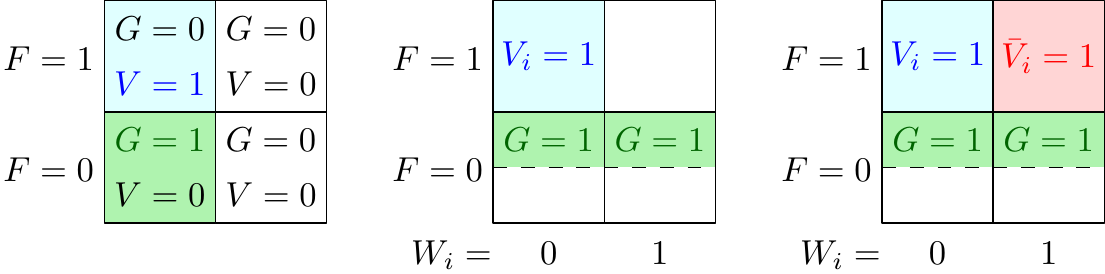}
\par\end{centering}
\caption{\label{fig:flip}Left: An illustration for $\mathrm{FLIP}(F,G,V)$,
showing the $4$ possible combinations of $(F,G,V)$ which are equally
likely. Middle: An illustration for $F,G,V_{i},W_{i}$ satisfying
$\mathrm{SW}(W^{k},V^{k},\bar{V}^{k},F)$. The x-axis is $W_{i}$,
which is independent of $F$ (the y-axis). Since $\mathrm{FLIP}(F,G,V_{i})$,
we can assume $G|F\sim\mathrm{Bern}((1-F)/2)$. Since $I(W_{i};F,G)=0$,
we have $G|\{(W_{i},F)=(0,0)\}\sim\mathrm{Bern}(1/2)$ and $G|\{(W_{i},F)=(1,0)\}\sim\mathrm{Bern}(1/2)$
(represented by dividing the cells $(W_{i},F)=(0,0)$ and $(1,0)$
into halves in the figure). Since $\mathrm{FLIP}(F,G,V_{i})$ and
$H(V_{i}|W_{i},F)=0$, we can assume $V_{i}=1$ if and only if $(W_{i},F)=(0,1)$
(another choice is $(W_{i},F)=(1,1)$). Right: An illustration for
$F,G,V_{i},\bar{V}_{i},W_{i}$ in $\mathrm{SW}(W^{k},V^{k},\bar{V}^{k},F)$.
We cannot have $\bar{V}_{i}=1$ if and only if $(W_{i},F)=(0,1)$
since $I(V_{i};\bar{V}_{i}|W_{i})=0$. Hence $\bar{V}_{i}=1$ if and
only if $(W_{i},F)=(1,1)$. The purpose of $G$ is to force $V_{i}$
and $\bar{V}_{i}$ to be $1$ for the same value of $F$.}
\end{figure}

Consider the distribution $F\sim\mathrm{Bern}(1/2)$, $G_{1}|F\sim\mathrm{Bern}((1-F)/2)$,
$G_{2}|(F,G_{1})\sim\mathrm{Bern}(F/2)$ (i.e., the distribution $(F,G_{1},G_{2})\sim\mathrm{Unif}(\{(0,0,0),(0,1,0),(1,0,0),(1,0,1)\})$).
We can check for this joint distribution (up to relabeling) by
\begin{align*}
 & \!\!\!\!\mathrm{FLIP}(F,G_{1},G_{2}):\\
 & \exists U,Z_{1},Z_{2}:\,\mathrm{UNIF}_{4}(U)\,\wedge\,\mathrm{UNIF}_{2}(F)\\
 & \wedge\,H(F,G_{1},G_{2}|U)=I(G_{1};G_{2}|F)=0\\
 & \wedge\,\mathrm{UNIF}_{3}(Z_{1})\,\wedge\,I(Z_{1};G_{1})=H(U|G_{1},Z_{1})=0\\
 & \wedge\,\mathrm{UNIF}_{3}(Z_{2})\,\wedge\,I(Z_{2};G_{2})=H(U|G_{2},Z_{2})=0.
\end{align*}
This is because $\mathrm{UNIF}_{4}(U)$, $H(G_{1}|U)=0$, $\mathrm{UNIF}_{3}(Z_{1})$
and $I(Z_{1};G_{1})=H(U|G_{1},Z_{1})=0$ implies $G_{1}$ either follows
$\mathrm{Unif}([4])$ or $\mathrm{Bern}(1/4)$ (up to relabeling).
Same for $G_{2}$. Since $I(G_{1};G_{2}|F)=0$, it is impossible to
have $G_{1}\sim\mathrm{Unif}([4])$ or $G_{2}\sim\mathrm{Unif}([4])$.
Hence we have the desired distribution. Refer to Figure \ref{fig:flip}
for an illustration.

\begin{figure}
\begin{centering}
\includegraphics[scale=0.9]{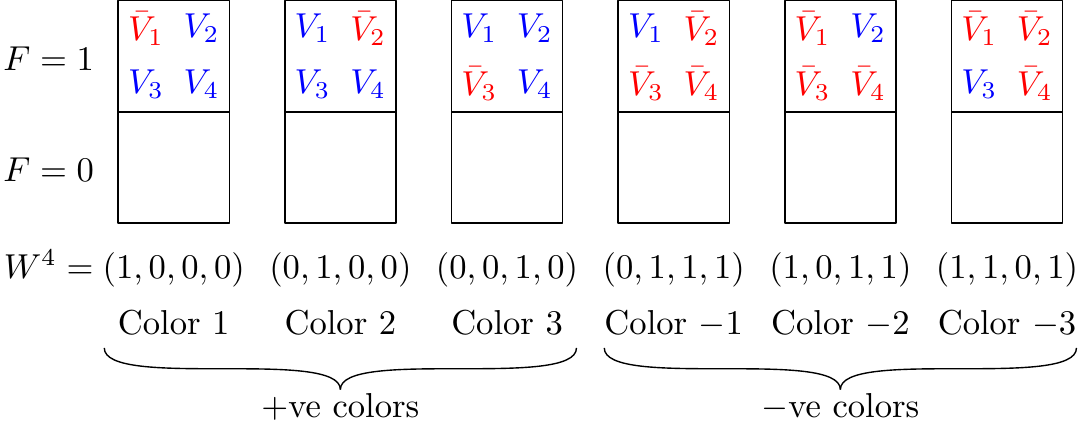}
\par\end{centering}
\caption{\label{fig:col1}The table of the values of $V^{k},\bar{V}^{k}$ (which
are functions of $W^{k},F$) for different values of $W^{k},F$ when
$k=4$. The x-axis denotes different values of $W^{4}$ (there are
$6$ possible combinations according to the constraints in $\mathrm{COL}$,
which are referred as colors), whereas the y-axis denotes the two
values of $F\in\{0,1\}$. If $V_{i}=1$ for a combination of $W^{k}$
and $F$, the corresponding cell in the table is marked \textquotedblleft$V_{i}$\textquotedblright{}
in blue (similar for $\bar{V}_{i}$ in red). This table can be obtained
from the relation $V_{i}=(1-W_{i})F$, $\bar{V}_{i}=W_{i}F$.}
\end{figure}

We introduce a construction called \emph{switch}: for $k\ge4$,
\begin{align}
 & \!\!\!\!\mathrm{SW}(W^{k},V^{k},\bar{V}^{k},F):\nonumber \\
 & \exists G:\,I(W^{k};F,G)=0\nonumber \\
 & \wedge\,\bigwedge_{i\in[k]}\big(\mathrm{UNIF}_{2}(W_{i})\nonumber \\
 & \;\;\wedge\,H(V_{i},\bar{V}_{i}|W_{i},F)=I(V_{i};\bar{V}_{i}|W_{i})=0\nonumber \\
 & \;\;\wedge\,\mathrm{FLIP}(F,G,V_{i})\,\wedge\,\mathrm{FLIP}(F,G,\bar{V}_{i})\big).\label{eq:sw}
\end{align}
We can regard $W^{k}\in\{0,1\}^{k}$ as a sequence of Boolean-valued
switches. Refer to Figures \ref{fig:flip} and \ref{fig:col1} for
an illustration.

Since $\mathrm{FLIP}(F,G,V_{i})$, we assume $F\sim\mathrm{Bern}(1/2)$,
$G|F\sim\mathrm{Bern}((1-F)/2)$. Since $\mathrm{UNIF}_{2}(W_{i})$,
$\mathrm{UNIF}_{2}(F)$ and $I(W_{i};F)=0$, we have $(W_{i},F)\sim\mathrm{Unif}(\{0,1\}^{2})$.
Since $\mathrm{FLIP}(F,G,V_{i})$, we can assume $V_{i}|F\sim\mathrm{Bern}(F/2)$.
Combining this with $H(V_{i}|W_{i},F)=0$, we either have $V_{i}=W_{i}F$
or $V_{i}=(1-W_{i})F$. Assume $V_{i}=(1-W_{i})F$ without loss of
generality. Similarly, we either have $\bar{V}_{i}=W_{i}F$ or $\bar{V}_{i}=(1-W_{i})F$.
The latter is impossible since $I(V_{i};\bar{V}_{i}|W_{i})=0$. Hence
we have $V_{i}=(1-W_{i})F$, $\bar{V}_{i}=W_{i}F$.

Since $I(W^{k};F)=0$, the conditional distribution of $F$ given
$W^{k}=w^{k}$ is $\mathrm{Bern}(1/2)$. Using $V_{i}=(1-W_{i})F$,
$\bar{V}_{i}=W_{i}F$, for $S,\bar{S}\subseteq[k]$,
\begin{align*}
H(F|V_{S},\bar{V}_{\bar{S}},W^{k}=w^{k}) & =\mathrm{sat}(w^{k},S,\bar{S}),
\end{align*}
where
\[
\mathrm{sat}(w^{k},S,\bar{S}):=\left(\prod_{i\in S}w_{i}\right)\left(\prod_{i\in\bar{S}}(1-w_{i})\right)\in\{0,1\},
\]
which is $1$ if and only if $w_{i}=1$ for all $i\in S$, and $w_{i}=0$
for all $i\in\bar{S}$. We have
\[
H(F|V_{S},\bar{V}_{\bar{S}},W^{k})=\mathbf{E}[\mathrm{sat}(W^{k},S,\bar{S})].
\]
In particular, for $w^{k}\in\{0,1\}^{k}$,
\[
H(F|V_{\{i:w_{i}=1\}},\bar{V}_{\{i:w_{i}=0\}},W^{k})=\mathbf{P}(W^{k}=w^{k}).
\]
There are a total of $2^{k}$ choices of $W^{k}$. We reduce the number
of choices by restricting that for any $w^{k}$ with $p_{W^{k}}(w^{k})>0$,
there exists $j\in[k-1]$ such that either $w_{i}=\mathbf{1}\{i=j\}$
for $i\in[k]$ (in this case, call the \emph{color} of $w^{k}$ to
be $j$, a positive color), or $w_{i}=\mathbf{1}\{i\neq j\}$ for
$i\in[k]$ (in this case, call the color of $w^{k}$ to be $-j$,
a negative color). Let the set of such $w^{k}$ be $T_{k}\subseteq\{0,1\}^{k}$.
There are $|T_{k}|=2(k-1)$ choices of colors. Let $\mathrm{col}:T_{k}\to[-(k-1)..-1]\cup[1..k-1]$
such that $\mathrm{col}(w^{k})$ is the color of $w^{k}$, and write
$\mathrm{sgncol}(w^{k}):=\mathrm{sgn}(\mathrm{col}(w^{k}))$ for the
sign of the color. Note that all $w^{k}$ of positive color has $w_{k}=0$,
and all $w^{k}$ of negative color has $w_{k}=1$. Define the coloring
AEIP as
\begin{align*}
 & \!\!\!\!\mathrm{COL}(W^{k},V^{k},\bar{V}^{k},F):\\
 & \mathrm{SW}(W^{k},V^{k},\bar{V}^{k},F)\\
 & \wedge\,\bigwedge_{w^{k}\in\{0,1\}^{k}\backslash T_{k}}\big(H(F|V_{\{i:w_{i}=1\}},\bar{V}_{\{i:w_{i}=0\}},W^{k})=0\big).
\end{align*}
Refer to Figure \ref{fig:col1} for an illustration. Note that the
``$\bigwedge$'' is a conjunction of finitely many choices of $w^{k}$
(that does not depend on the random variables), and hence the resultant
predicate is still an AEIP. The last condition ensures that $\mathbf{P}(W^{k}=w^{k})=0$
for $w^{k}\notin T_{k}$, and $W^{k}\in T_{k}$ with probability $1$.

\medskip{}

\subsection{Color Constraints on Edges}

We now consider applying the colors to vertices. Let $X$ be a random
variable representing a uniformly chosen random vertex. We assign
a color to each vertex (i.e., the color $W^{k}$ is a function of
$X$), as given by the following AEIP:
\begin{align*}
 & \!\!\!\!\mathrm{COLD}(X,W^{k},V^{k},\bar{V}^{k},F):\\
 & \mathrm{COL}(W^{k},V^{k},\bar{V}^{k},F)\\
 & \wedge\,H(W^{k}|X)=I(V^{k},\bar{V}^{k},F;X|W^{k})=0.
\end{align*}
Let $W^{k}=w^{k}(X)$ be a function of $X$. Let $E$ be a random
variable where $H(E|X)=0$ and $p_{X|E=e}$ is uniform over $l$ values
for any $e$. Regard $E$ as groups of vertices of size $l$. Letting
$p_{X|E=e}=\mathrm{Unif}\{x(e,1),\ldots,x(e,l)\}$, we have the conditional
distribution
\begin{align*}
 & (F,V_{S},\bar{V}_{\bar{S}})|\{E=e\}\\
 & \sim l^{-1}\sum_{i=1}^{l}p_{F,V_{S},\bar{V}_{\bar{S}}|W^{k}=w^{k}(x(e,i))},
\end{align*}
which is due to the conditional independence $I(V^{k},\bar{V}^{k},F;X|W^{k})=0$.

We can enforce certain rules on the colors of vertices in a group.
For example, to enforce that for each group, all of the vertices must
have positive color (recall that $w^{k}$ has positive color if and
only if $w_{k}=0$), we can use the condition
\[
H(F|V_{\{k\}},\bar{V}_{\emptyset},E)=0.
\]
This is because $V_{k}=(1-W_{k})F$, so if $F$ is a function of $V_{k}$
and $E$, we must have $W_{k}=0$ with probability $1$. Note that
we will not actually enforce this.

\begin{figure}
\begin{centering}
\includegraphics[scale=0.9]{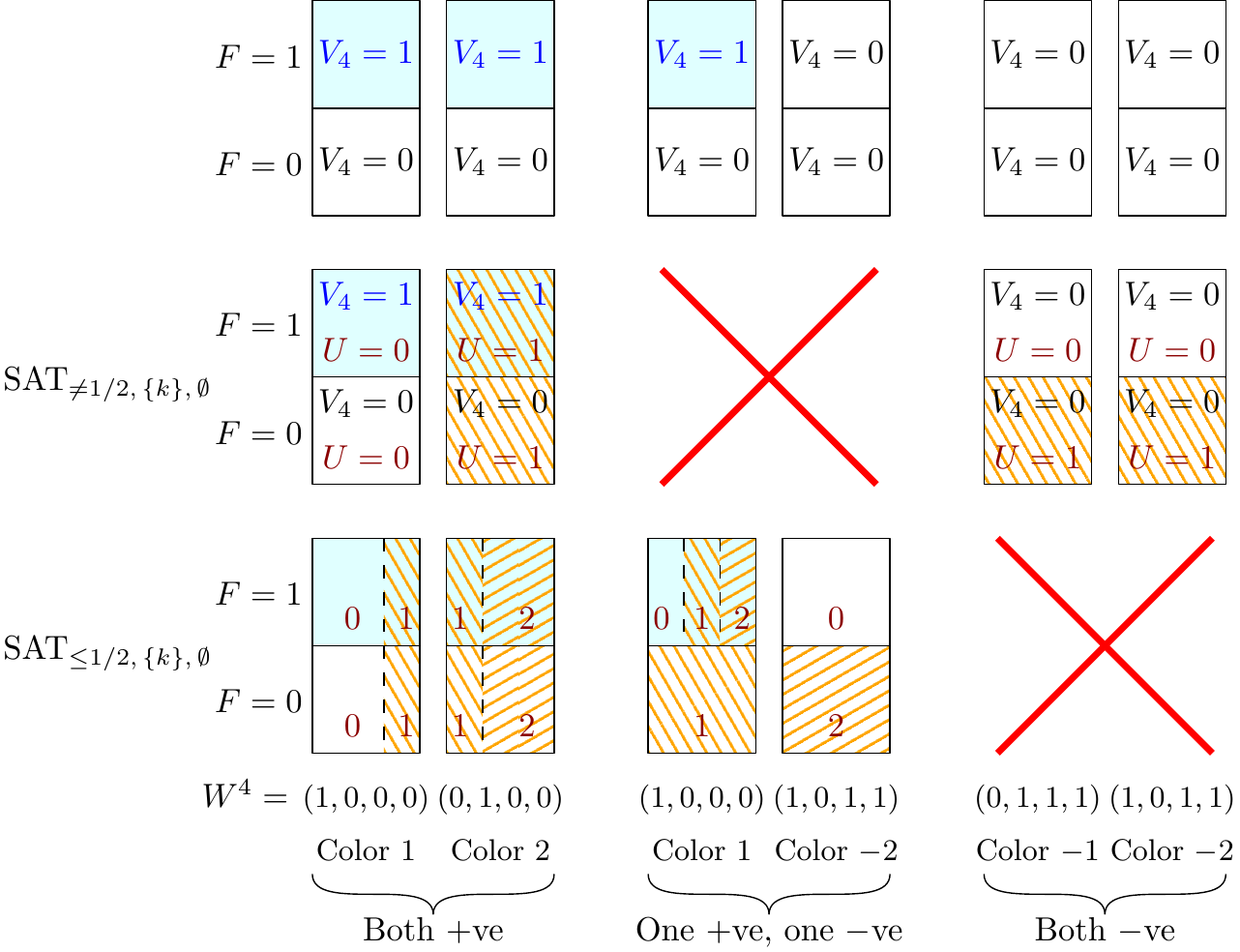}
\par\end{centering}
\caption{\label{fig:col2}An illustration of $\mathrm{SAT}_{\protect\neq1/2,\,\{k\},\,\emptyset}$
\eqref{eq:sat_n12} (second row) and $\mathrm{SAT}_{\le1/2,\,\{k\},\,\emptyset}$
(third row) for $k=4$. Refer to Figure \ref{fig:col1} for the meaning
of the axes. For the first case (first column) where the two sequences
$W^{k}$ on the two vertices are both positive, we have $F=V_{4}$,
and hence there exists $U\sim\mathrm{Bern}(1/2)$ (independent of
$V_{4}$) where $F$ is a function of $(V_{4},U)$, and $\mathrm{SAT}_{\protect\neq1/2,\,\{k\},\,\emptyset}$
is true (the hatch patterns in the figure represent different values
of $U$). For the second case (second column) where the two sequences
$W^{k}$ have different signs, we have $F|\{V_{4}=0\}\sim\mathrm{Bern}(1/3)$
and $F=1$ when $V_{4}=1$, and hence there does not exist such $U\sim\mathrm{Bern}(1/2)$,
and $\mathrm{SAT}_{\protect\neq1/2,\,\{k\},\,\emptyset}$ is false.
For the third case (third column) where the two sequences $W^{k}$
are both negative, we have $V_{4}=0$ independent of $F\sim\mathrm{Bern}(1/2)$,
and hence there exists such $U\sim\mathrm{Bern}(1/2)$ (take $U=F$),
and $\mathrm{SAT}_{\protect\neq1/2,\,\{k\},\,\emptyset}$ is true.
The cases for $\mathrm{SAT}_{\le1/2,\,\{k\},\,\emptyset}$ are similar.}
\end{figure}

Consider the case $l=2$, and we want to enforce that each group contains
two vertices of the same signs of colors. If all of the vertices have
negative color, since $V_{k}=(1-W_{k})F=0$, there exists $U\sim\mathrm{Bern}(1/2)$
(independent of $E,V_{k}$) such that $H(F|V_{k},E,U)=0$ (simply
take $U=F$). If exactly one of the two vertices has negative color,
since $V_{k}=(1-W_{k})F$, we have $V_{k}=0$ with probability $1/2$,
and $V_{k}=F$ with probability $1/2$. We have $F|\{V_{k}=0,E=e\}\sim\mathrm{Bern}(1/3)$,
and hence there does not exist $U\sim\mathrm{Bern}(1/2)$ (independent
of $E,V_{k}$) such that $H(F|V_{k},E,U)=0$. Hence the condition
that each group contains two vertices of the same signs of colors
can be checked using the AEIP $\mathrm{SAT}_{\neq1/2,\,\{k\},\,\emptyset}(E,W^{k},V^{k},\bar{V}^{k},F)$,
where
\begin{align}
 & \!\!\!\!\mathrm{SAT}_{\neq1/2,\,S,\,\bar{S}}(E,W^{k},V^{k},\bar{V}^{k},F):\nonumber \\
 & \exists U:\,\mathrm{UNIF}_{2}(U)\,\wedge\,I(U;E,V_{S},\bar{V}_{\bar{S}})=0\nonumber \\
 & \wedge\,H(F|V_{S},\bar{V}_{\bar{S}},E,U)=0.\label{eq:sat_n12}
\end{align}
This means that for each group, the number of vertices $w^{k}$ with
$\mathrm{sat}(w^{k},S,\bar{S})=1$ cannot be exactly one (out of $l=2$
vertices). Refer to Figure \ref{fig:col2} for an illustration.

Now consider $l=2$ and we want to enforce that the number of vertices
$w^{k}$ with $\mathrm{sat}(w^{k},S,\bar{S})=1$ in each group is
at most one. Using a similar argument, we have
\begin{align}
 & \!\!\!\!\mathrm{SAT}_{\le1/2,\,S,\,\bar{S}}(E,W^{k},V^{k},\bar{V}^{k},F):\nonumber \\
 & \exists U:\,\mathrm{UNIF}_{3}(U)\,\wedge\,I(U;E,V_{S},\bar{V}_{\bar{S}})=0\nonumber \\
 & \wedge\,H(F|V_{S},\bar{V}_{\bar{S}},E,U)=0.\label{eq:sat_le12}
\end{align}
Note that a function of a random variable following $\mathrm{Unif}([3])$
can be $\mathrm{Bern}(1/3)$, but cannot be $\mathrm{Bern}(1/2)$. 

Now consider $l=4$ and we want to enforce that the number of vertices
$w^{k}$ with $\mathrm{sat}(w^{k},S,\bar{S})=1$ in each group is
at most three. Note that $F=1$ if $V_{S}\neq0$ or $\bar{V}_{\bar{S}}\neq0$,
and $F|\{V_{S}=0,\bar{V}_{\bar{S}}=0,E=e\}\sim\mathrm{Bern}(a_{e}/(4+a_{e}))$,
where $a_{e}$ is the number of vertices $w^{k}$ with $\mathrm{sat}(w^{k},S,\bar{S})=1$
in the group $E=e$. Using a similar argument, we have
\begin{align}
 & \!\!\!\!\mathrm{SAT}_{\le3/4,\,S,\,\bar{S}}(E,W^{k},V^{k},\bar{V}^{k},F):\nonumber \\
 & \exists U:\,\mathrm{UNIF}_{105}(U)\,\wedge\,I(U;E,V_{S},\bar{V}_{\bar{S}})=0\nonumber \\
 & \wedge\,H(F|V_{S},\bar{V}_{\bar{S}},E,U)=0.\label{eq:sat_le34}
\end{align}
Note that a function of a random variable following $\mathrm{Unif}([105])$
can be $\mathrm{Bern}(1/5)$, $\mathrm{Bern}(1/3)$ or $\mathrm{Bern}(3/7)$,
but cannot be $\mathrm{Bern}(1/2)$.

\medskip{}

\subsection{Construction of Colored Tori}

For $k\ge1$, the \emph{colored tori} is defined as
\begin{align*}
 & \!\!\!\!\mathrm{CTORI}(X^{2},Y^{2},W^{k},V^{k},\bar{V}^{k},F):\\
 & \mathrm{TORI}(X^{2},Y^{2})\\
 & \wedge\,\mathrm{COLD}((X^{2},Y^{2}),W^{k},V^{k},\bar{V}^{k},F).
\end{align*}
Write $\mathcal{R}$ for the support of $p_{X^{2},Y^{2}}$. Since
$W_{i}$ is a function of $X^{2},Y^{2}$, we denote that function
as $w_{i}(x_{1},x_{2},y_{1},y_{2})=w_{i}(x^{2},y^{2})$, and let $w^{k}(x^{2},y^{2})=\{w_{i}(x^{2},y^{2})\}_{i\in[k]}$
for $(x^{2},y^{2})\in\mathcal{R}$. Each vertex of the tori is assigned
one of $2(k-1)$ colors.

We remark that since \eqref{eq:sw} enforces $\mathrm{UNIF}_{2}(W_{i})$,
the number of vertices with $w_{i}(x^{2},y^{2})=0$ must be equal
to the number of vertices with $w_{i}(x^{2},y^{2})=1$. This is satisfied
if for each $j\in[k-1]$, the number of vertices with color $j$ is
equal to the number of vertices with color $-j$. This will be addressed
in Section \ref{subsec:reduction}.

Conditioned on $(X_{1},X_{2},Y_{1})=(x_{1},x_{2},y_{1})$, the distribution
of $(X^{2},Y^{2})$ is uniform over two values. The tuple $(x_{1},x_{2},y_{1})$
corresponds to an edge in the tori, which we call a \emph{vertical}
edge (since the two vertices have different vertical coordinates $y_{2}$).
We can check whether the two vertices have the same signs of colors
using \eqref{eq:sat_n12}:
\[
\mathrm{SAT}_{\neq1/2,\,\{k\},\,\emptyset}((X_{1},X_{2},Y_{1}),W^{k},V^{k},\bar{V}^{k},F).
\]
Similar for conditioning on $(X_{1},X_{2},Y_{2})$ (vertical edge),
$(X_{1},Y_{1},Y_{2})$ (horizontal edge) or $(X_{2},Y_{1},Y_{2})$
(horizontal edge). Hence we can use AEIPs to force the signs of colors
to be constant within the same torus, and we call it the sign of the
torus. Note that different tori within the collection of tori may
have different signs.

\begin{figure}
\begin{centering}
\includegraphics[scale=0.8]{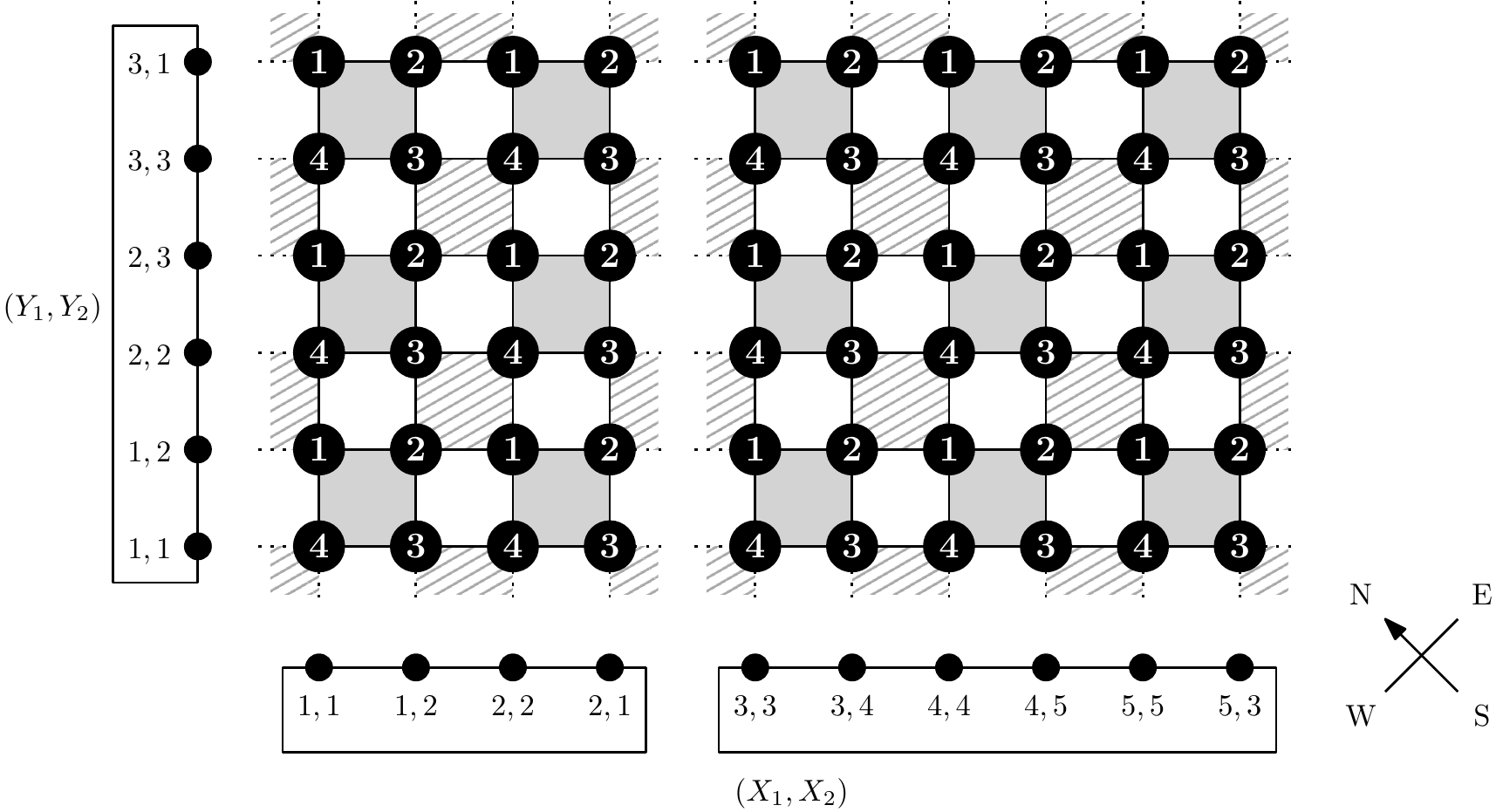}
\par\end{centering}
\caption{\label{fig:tori2}The oriented tori. The label on each vertex is its
group. Solid grey squares are type $11$ faces, whereas hatched grey
squares are type $22$ faces. }
\end{figure}

We now require that $k-1\ge8$ is a multiple of $4$. Divide the colors
into $4$ groups according to the remainder modulo $4$ of their absolute
values (i.e., color $j$ is in group $|j|\;\mathrm{mod}\,4$, where
we assume $a\;\mathrm{mod}\,4\in[4]$ instead of $[0..3]$).  We
enforce that each horizontal edge either connect a group $1$ vertex
and a group $2$ vertex (the group of a vertex is the group of its
color), or connect a group $3$ vertex and a group $4$ vertex. Also,
each vertical edge either connect a group $1$ vertex and a group
$4$ vertex, or connect a group $2$ vertex and a group $3$ vertex.
Note that $\mathrm{sat}(w^{k},\emptyset,[3..k])=\mathrm{sat}(\tilde{w}^{k},\emptyset,[3..k])=1$
if and only if $\mathrm{col}(w^{k}),\mathrm{col}(\tilde{w}^{k})\in\{1,2\}$
(so they are both group $1$ or $2$), and hence this should be forbidden
in a vertical edge. Altogether, we can define the \emph{oriented tori}
using \eqref{eq:sat_le12} as 
\begin{align}
 & \!\!\!\!\mathrm{OTORI}(X^{2},Y^{2},W^{k},V^{k},\bar{V}^{k},F):\nonumber \\
 & \mathrm{CTORI}(X^{2},Y^{2},W^{k},V^{k},\bar{V}^{k},F)\nonumber \\
 & \wedge\,\mathrm{SAT}_{\neq1/2,\,\{k\},\,\emptyset}((X_{1},X_{2},Y_{1}),W^{k},V^{k},\bar{V}^{k},F)\nonumber \\
 & \wedge\,\mathrm{SAT}_{\neq1/2,\,\{k\},\,\emptyset}((X_{1},X_{2},Y_{2}),W^{k},V^{k},\bar{V}^{k},F)\nonumber \\
 & \wedge\,\mathrm{SAT}_{\neq1/2,\,\{k\},\,\emptyset}((X_{1},Y_{1},Y_{2}),W^{k},V^{k},\bar{V}^{k},F)\nonumber \\
 & \wedge\,\mathrm{SAT}_{\neq1/2,\,\{k\},\,\emptyset}((X_{2},Y_{1},Y_{2}),W^{k},V^{k},\bar{V}^{k},F)\nonumber \\
 & \wedge\,\bigwedge_{j_{1},j_{2}\in[1..k-1]:\,\{j_{1}\,\mathrm{mod}\,4,\,j_{2}\,\mathrm{mod}\,4\}\notin\{\{1,4\},\{2,3\}\}}\big(\nonumber \\
 & \;\;\;\;\mathrm{SAT}_{\le1/2,\,\emptyset,\,[k]\backslash\{j_{1},j_{2}\}}((X_{1},X_{2},Y_{1}),W^{k},V^{k},\bar{V}^{k},F)\nonumber \\
 & \;\;\;\;\wedge\,\mathrm{SAT}_{\le1/2,\,\emptyset,\,[k]\backslash\{j_{1},j_{2}\}}((X_{1},X_{2},Y_{2}),W^{k},V^{k},\bar{V}^{k},F)\nonumber \\
 & \;\;\;\;\wedge\,\mathrm{SAT}_{\le1/2,\,[k]\backslash\{j_{1},j_{2}\},\,\emptyset}((X_{1},X_{2},Y_{1}),W^{k},V^{k},\bar{V}^{k},F)\nonumber \\
 & \;\;\;\;\wedge\,\mathrm{SAT}_{\le1/2,\,[k]\backslash\{j_{1},j_{2}\},\,\emptyset}((X_{1},X_{2},Y_{2}),W^{k},V^{k},\bar{V}^{k},F)\big)\nonumber \\
 & \wedge\,\bigwedge_{j_{1},j_{2}\in[1..k-1]:\,\{j_{1}\,\mathrm{mod}\,4,\,j_{2}\,\mathrm{mod}\,4\}\notin\{\{1,2\},\{3,4\}\}}\big(\nonumber \\
 & \;\;\;\;\mathrm{SAT}_{\le1/2,\,\emptyset,\,[k]\backslash\{j_{1},j_{2}\}}((X_{1},Y_{1},Y_{2}),W^{k},V^{k},\bar{V}^{k},F)\nonumber \\
 & \;\;\;\;\wedge\,\mathrm{SAT}_{\le1/2,\,\emptyset,\,[k]\backslash\{j_{1},j_{2}\}}((X_{2},Y_{1},Y_{2}),W^{k},V^{k},\bar{V}^{k},F)\nonumber \\
 & \;\;\;\;\wedge\,\mathrm{SAT}_{\le1/2,\,[k]\backslash\{j_{1},j_{2}\},\,\emptyset}((X_{1},Y_{1},Y_{2}),W^{k},V^{k},\bar{V}^{k},F)\nonumber \\
 & \;\;\;\;\wedge\,\mathrm{SAT}_{\le1/2,\,[k]\backslash\{j_{1},j_{2}\},\,\emptyset}((X_{2},Y_{1},Y_{2}),W^{k},V^{k},\bar{V}^{k},F)\big),\label{eq:otori}
\end{align}
where $a\,\mathrm{mod}\,b$ is defined to take values in $[b]$ (instead
of $[0..b-1]$). Refer to Figure \ref{fig:tori2} for an illustration.

Conditioned on $(X_{1},Y_{1})=(x_{1},y_{1})$, the distribution of
$(X^{2},Y^{2})$ is uniform over $4$ values. We call these $4$ vertices
a \emph{type} $11$ \emph{face}. Due to the aforementioned constraint
on the groups, each face has vertices of all $4$ groups. Regarding
the group $1,2,3,4$ vertices of a type $11$ face as the north (N),
east (E), south (S) and west (W) directions respectively, we can fix
the orientation of the torus (see Figure \ref{fig:tori2}). Note that
the N, E, S, W directions are rotated $45^{\circ}$ compared to the
horizontal/vertical directions of the edges.

Conditioned on $(X_{2},Y_{2})=(x_{2},y_{2})$, the distribution of
$(X^{2},Y^{2})$ is uniform over $4$ values. We call these $4$ vertices
a \emph{type} $22$ \emph{face}. Note that the group $1,2,3,4$ vertices
of a type $22$ face are pointing to the S, W, N and E directions
respectively.

We use type $11$ and type $22$ faces as our grid (ignore type $12$
and type $21$ faces). Each type $11$ face has $4$ neighboring type
$22$ faces in the N, E, S, W directions respectively. The grid is
rotated $45^{\circ}$ compared to the horizontal/vertical directions
of the edges.

\medskip{}

\subsection{Reduction from the Periodic Tiling Problem\label{subsec:reduction}}

We now show a reduction from the periodic tiling problem \cite{gurevich1972remarks}.
A tiles can be represented as an integer sequence of length $4$,
with entries representing the colors of the N,E,S,W edges respectively.
Assume there are $(k-1)/4$ different colors among the edges of tiles
(\emph{tile colors}), and let the set of tiles be $\mathcal{C}\subseteq[(k-1)/4]^{4}$.
Assume $k-1\ge8$ without loss of generality. Recall that the number
of colors of vertices of the tori is $2(k-1)$ (\emph{vertex colors}).
Each tile color $j\in[(k-1)/4]$ corresponds to $8$ vertex colors:
$4j-3,\,4j-2,\,4j-1,\,4j,\,-(4j-3),\,-(4j-2),\,-(4j-1),\,-4j$, one
per combination of sign and group. To enforce that each type $11$
face has colors belonging to one of the tiles in $\mathcal{C}$, the
set of absolute values of colors of the $4$ vertices must be in the
form 
\begin{align*}
 & \{4c_{1}-3,\,4c_{2}-2,\,4c_{3}-1,\,4c_{4}\}
\end{align*}
for some $c\in\mathcal{C}$ (e.g. $4c_{1}-3$ is a positive group
$1$ color, which appears in the north vertex of a type $11$ face
in a positive torus, so this means the north vertex has a vertex color
that corresponds to the tile color $c_{1}$). Rotations and reflections
are disallowed since the vertex colors contain group (and hence orientation)
information. Write $\mathcal{C}_{11}:=\{\{4c_{1}-3,\,4c_{2}-2,\,4c_{3}-1,\,4c_{4}\}:\,c\in\mathcal{C}\}$.
Similarly, to enforce that each type $22$ face has colors belonging
to $\mathcal{C}$, the set of absolute values of colors must be in
the form 
\begin{align*}
 & \{4c_{1}-1,\,4c_{2},\,4c_{3}-3,\,4c_{4}-2\}
\end{align*}
for some $c\in\mathcal{C}$. Define $\mathcal{C}_{22}$ similarly.
These constraints can be enforced in a similar manner as \eqref{eq:otori}
using \eqref{eq:sat_le34}. The constraint that touching edges of
adjacent tiles match is automatically enforced since two adjacent
faces share a vertex. Therefore, the final AEIP is given by
\begin{align*}
 & \!\!\!\!\mathrm{TTORI}_{\mathcal{C}}:\\
 & \exists X^{2},Y^{2},W^{k},V^{k},\bar{V}^{k},F:\\
 & \mathrm{OTORI}(X^{2},Y^{2},W^{k},V^{k},\bar{V}^{k},F)\\
 & \wedge\,\bigwedge_{j_{1},\ldots,j_{4}\in[1..k-1]:\,j_{i}\,\mathrm{mod}\,4\;\text{distinct},\,\{j_{1},\ldots,j_{4}\}\notin\mathcal{C}_{11}}\big(\\
 & \;\;\;\;\mathrm{SAT}_{\le3/4,\,\emptyset,\,[k]\backslash\{j_{1},\ldots,j_{4}\}}((X_{1},Y_{1}),W^{k},V^{k},\bar{V}^{k},F)\\
 & \;\;\;\;\wedge\,\mathrm{SAT}_{\le3/4,\,[k]\backslash\{j_{1},\ldots,j_{4}\},\,\emptyset}((X_{1},Y_{1}),W^{k},V^{k},\bar{V}^{k},F)\big)\\
 & \wedge\,\bigwedge_{j_{1},\ldots,j_{4}\in[1..k-1]:\,j_{i}\,\mathrm{mod}\,4\;\text{distinct},\,\{j_{1},\ldots,j_{4}\}\notin\mathcal{C}_{22}}\big(\\
 & \;\;\;\;\mathrm{SAT}_{\le3/4,\,\emptyset,\,[k]\backslash\{j_{1},\ldots,j_{4}\}}((X_{2},Y_{2}),W^{k},V^{k},\bar{V}^{k},F)\\
 & \;\;\;\;\wedge\,\mathrm{SAT}_{\le3/4,\,[k]\backslash\{j_{1},\ldots,j_{4}\},\,\emptyset}((X_{2},Y_{2}),W^{k},V^{k},\bar{V}^{k},F)\big).
\end{align*}

\begin{figure}
\begin{centering}
\includegraphics[scale=0.6]{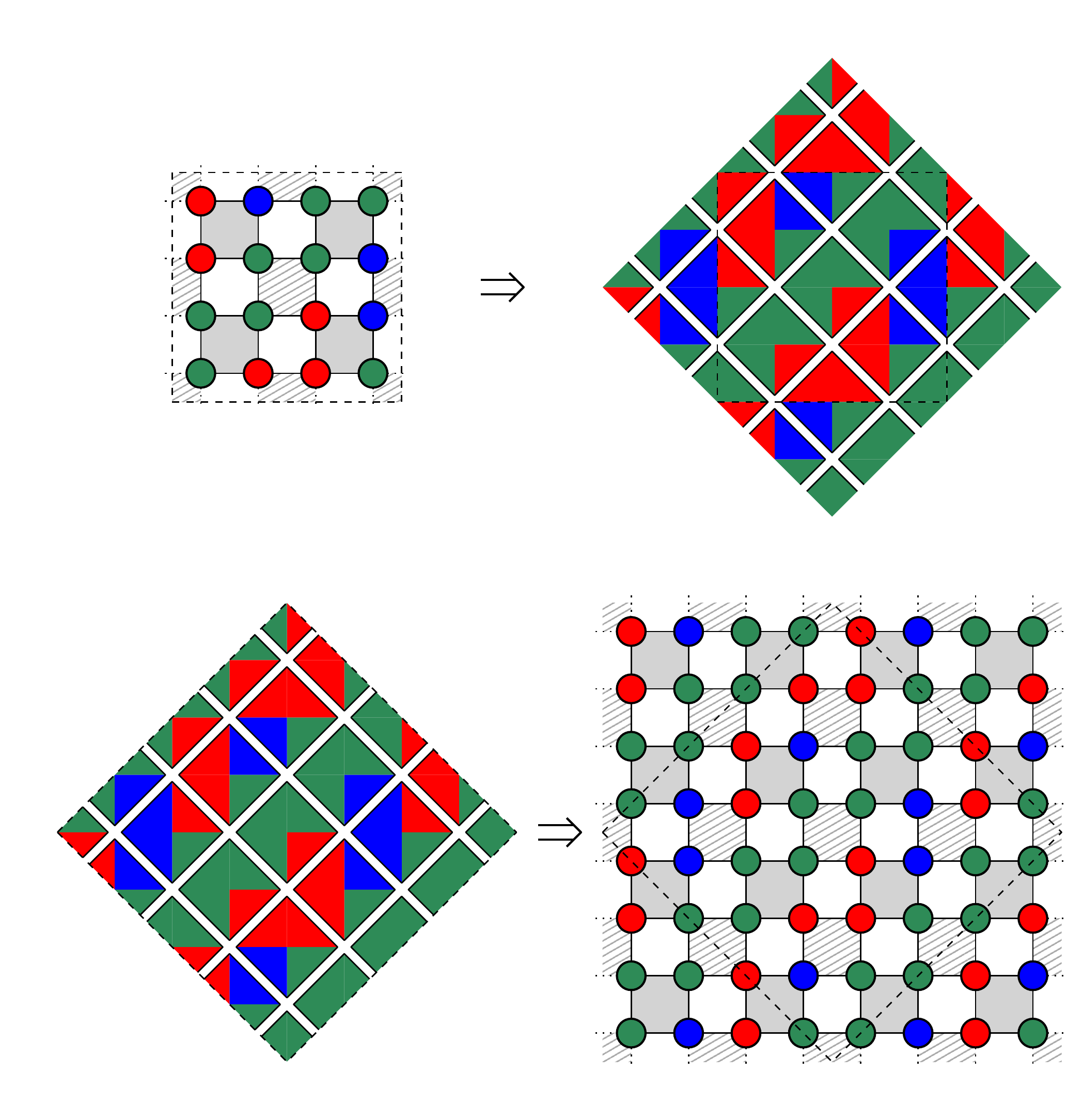}
\par\end{centering}
\caption{\label{fig:tori4}Top: Constructing a $4\times4$ periodic tiling
(right; the square tiles are rotated $45^{\circ}$; note that the
top left and the bottom right edges wrap around, and the top right
and the bottom left edges wrap around) from a $4\times4$ colored
torus (left). Bottom: Constructing a $8\times8$ colored torus (right)
from a $4\times4$ periodic tiling (left). Note that to ensure the
number of vertices with color $j$ is equal to the number of vertices
with color $-j$ (so $\mathrm{UNIF}_{2}(W_{i})$ in \eqref{eq:sw}
is satisfied), we have to create two copies of the $8\times8$ colored
torus on the right, one with positive colors and another with negative
colors.}
\end{figure}

Finally, we show that $\mathrm{TTORI}_{\mathcal{C}}$ is true if and
only if the tiles in $\mathcal{C}$ can tile the plane periodically.
For the ``only if'' direction, assume $\mathrm{TTORI}_{\mathcal{C}}$
is true. Consider any one of the tori, and assume the set of values
of $(X^{2},Y^{2})$ over that torus is
\[
\Big(\bigcup_{i\in\mathbb{Z}_{a}}\{(i,i),(i,i+1)\}\Big)\times\Big(\bigcup_{j\in\mathbb{Z}_{b}}\{(j,j),(j,j+1)\}\Big)
\]
without loss of generality, where $\mathbb{Z}_{a}=\mathbb{Z}/a\mathbb{Z}$
is the cyclic group of order $a$. For simplicity, we relabel the
elements of the above set by the mapping $(i_{1},i_{2},j_{1},j_{2})\mapsto(i_{1}+i_{2},j_{1}+j_{2})\in\mathbb{Z}_{2a}\times\mathbb{Z}_{2b}$,
so we have a coloring of the torus $\mathbb{Z}_{2a}\times\mathbb{Z}_{2b}$.
Repeat this coloring $l/a$ times horizontally and $l/b$ times vertically,
where $l=\mathrm{lcm}(a,b)$, to obtain a coloring of the torus $\mathbb{Z}_{2l}\times\mathbb{Z}_{2l}$.
We call this set the \emph{colored torus}. Note that each type 11
face has a set of four vertices in the form $\{2i,2i+1\}\times\{2j,2j+1\}$,
and each type 22 face has a set of four vertices in the form $\{2i+1,2i+2\}\times\{2j+1,2j+2\}$.
Therefore each type 11 or 22 face (we call this an \emph{even} face)
has a set of four vertices in the form $\{i,i+1\}\times\{j,j+1\}$
where $i+j$ is even. We call this the even face at position $(i,j)\in\mathrm{even}(\mathbb{Z}_{2l}^{2}):=\{(i',j')\in\mathbb{Z}_{2l}^{2}:\,i'+j'\;\mathrm{even}\}$.
The even face at $(i,j)$ has neighbors (i.e., the other even faces
that shares one vertex with this even face) $(i\pm1,j\pm1)$. Define
the mapping $f:\mathbb{Z}_{2l}^{2}\to\mathrm{even}(\mathbb{Z}_{2l}^{2})$
from the \emph{tiling torus} $\mathbb{Z}_{2l}^{2}$ to the even faces
of the colored torus, by $f(u,v):=(u+v,\,u-v)$. Note that the even
face $f(u,v)$ has neighbors $f(u+1,v),f(u-1,v),f(u,v+1),f(u,v-1)$.
We now define a tiling over the tiling torus $\mathbb{Z}_{2l}^{2}$,
where each point $(u,v)\in\mathbb{Z}_{2l}^{2}$ corresponds to a square
tile, and the right edge of this tile has color which equals to the
color of the common vertex between even face $f(u,v)$ and $f(u+1,v)$
in the colored torus (this vertex is $(u+v+1,u-v+1)$), and similar
for the other edges. Hence we have a tiling of the plane with period
$2l$. Refer to the top figure in Figure \ref{fig:tori4}.

For the ``if'' direction, assume the tiles in $\mathcal{C}$ can
tile the plane periodically, and let the period be $a$ (i.e., it
tiles $\mathbb{Z}_{a}^{2}$). Using a similar argument as the ``only
if'' part, we have a coloring over $\mathbb{Z}_{2a}^{2}$. To ensure
that for each $j\in[k]$, the number of vertices with color $j$ is
equal to the number of vertices with color $-j$, so $\mathrm{UNIF}_{2}(W_{i})$
in \eqref{eq:sw} is satisfied, we use two copies of $\mathbb{Z}_{2a}^{2}$,
one with positive colors and the other with negative colors. The collection
of tori in $(X^{2},Y^{2})$ consists of these two tori. Refer to the
bottom figure in Figure \ref{fig:tori4}.

For the second part of Theorem \ref{thm:undecidable}, the undecidability
of the truth value of $\exists U^{l}\!:\mathbf{A}\mathbf{h}(U^{l})=\mathbf{b}$
follows from a reduction from $\exists U^{l}\!:\mathbf{A}\mathbf{h}(U^{l})\succeq\mathbf{b}$
since $\exists U^{l}\!:\mathbf{A}\mathbf{h}(U^{l})\succeq\mathbf{b}$
if and only if $\exists U^{l},V^{m}\!:\mathbf{A}\mathbf{h}(U^{l})-[H(V_{1}),\ldots,H(V_{m})]^{T}=\mathbf{b}$.
This completes the proof of Theorem \ref{thm:undecidable}.

\begin{figure}
\begin{centering}
\includegraphics[scale=0.8]{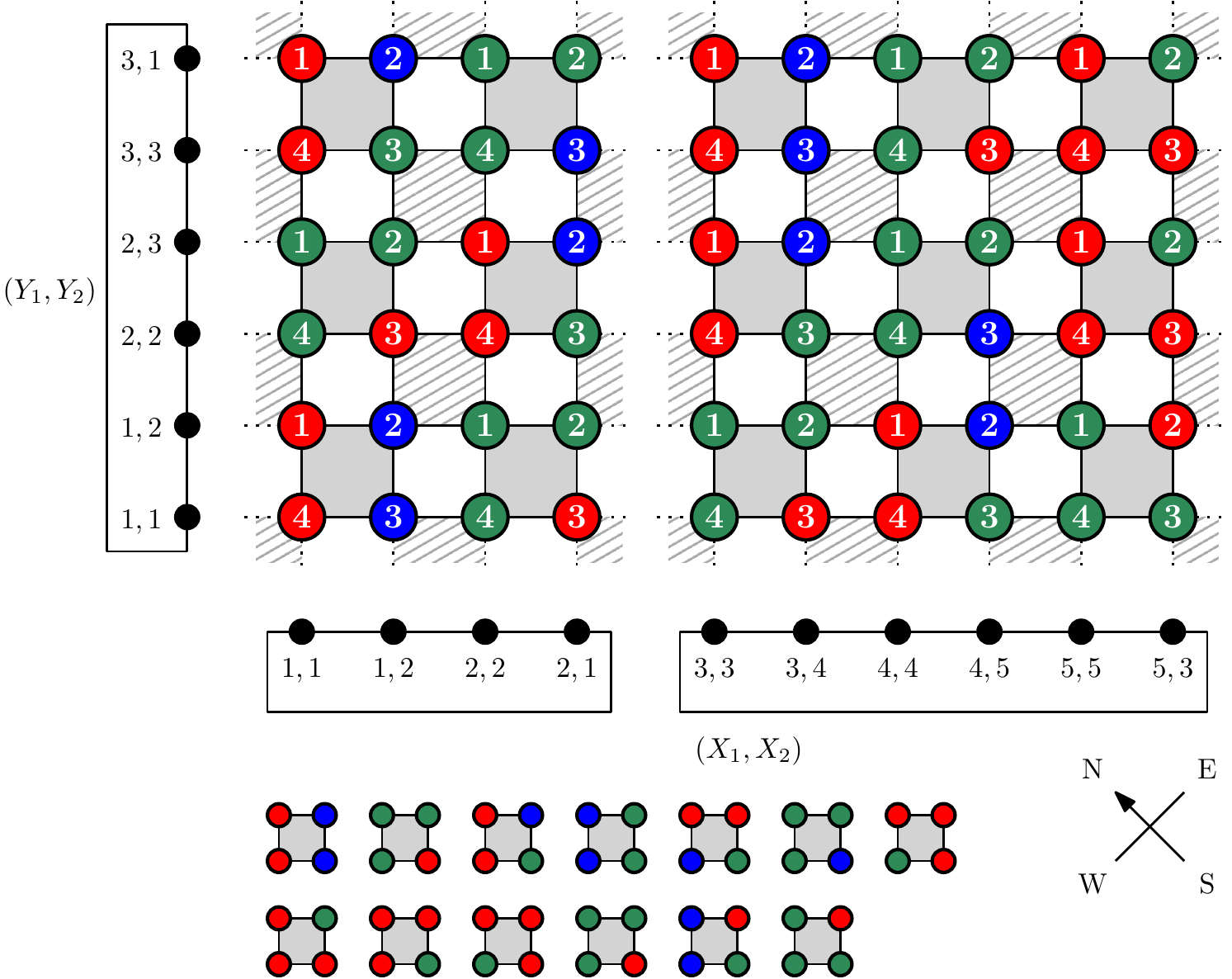}
\par\end{centering}
\caption{\label{fig:tori3}The tiled tori. Each vertex is colored red, green
or blue. Each (solid/hatched) grey square must belong to one of the
allowed tiles on the bottom.}
\end{figure}

\medskip{}

\section{Undecidability of Conditional Independence Implication with a Binary
Constraint\label{sec:ci}}

In this section, we show that the above proof also establish the following
result on the undecidability of the existence of random variables
satisfying some conditional independence relations and the constraint
that the first random variable is a binary uniform random variable. 

In the study of conditional independence implication, some authors
allow the three sets $A_{j},B_{j},C_{j}$ of random variables (in
the conditional independence relation $I(X_{A_{j}};X_{B_{j}}|X_{C_{j}})=0$)
to be non-disjoint \cite{dawid1979conditional,van1998informational,niepert2012logical},
whereas some authors require those sets to be disjoint \cite{pearl1987graphoids,studeny1997semigraphoids}
(call $I(X_{A_{j}};X_{B_{j}}|X_{C_{j}})=0$ a \emph{disjoint conditional
independence relation} in this case). The non-disjoint case allows
functional dependency in the form $H(X_{A_{j}}|X_{C_{j}})=0$ (by
letting $B_{j}=A_{j}$), whereas the disjoint case disallows this.
We will argue in Section \ref{subsec:disjoint_case} that the two
cases are Turing equivalent by showing a reduction from the non-disjoint
case to the disjoint case. This appears to be a new result, considering
\cite{studeny1997semigraphoids} (remark after Lemma 6) explicitly
disallow non-disjoint sets of random variables on the ground that
functional dependency is fundamentally different from conditional
independence relations (our result shows that this is not true, as
one can express functional dependency using conditional independence).

We first prove the non-disjoint case, and then prove the disjoint
case.

\subsection{Non-disjoint Case\label{subsec:nondisjoint}}

We first prove the following result that allows non-disjoint $A_{j},B_{j},C_{j}$.
\begin{thm}
\label{thm:ci_imp}The problem of deciding whether there exists random
variables $X^{n}$ such that $I(X_{A_{j}};X_{B_{j}}|X_{C_{j}})=0$
for all $j\in[m]$ and $X_{1}\sim\mathrm{Bern}(1/2)$, where $A_{j},B_{j},C_{j}\subseteq[n]$,
$n$ and $m$ are given, is undecidable.
\end{thm}
\begin{IEEEproof}
Note that throughout the proof of Theorem \ref{thm:undecidable},
we have only used conditional independence constraints in the form
$I(X;Y|Z)=0$ (note that $H(X|Z)=I(X;X|Z)$; this also includes $\mathrm{UNIF}(X)$),
and $\mathrm{UNIF}_{a}(Y)$ for $a\in\{2,3,4,105\}$. In this section,
we will use only conditional independence constraints and $\mathrm{UNIF}_{2}$,
and show that these constraints suffice. 

Using \eqref{eq:triple}, we can define the following AEIP that checks
whether $Y,Z$ are both uniform and have the same cardinality:
\begin{align*}
 & \!\!\!\!\mathrm{UNIF}_{=}(Y;Z):\,\\
 & \exists U^{3}:\,\mathrm{TRIPLE}(Y,U_{1},U_{2})\,\wedge\,\mathrm{TRIPLE}(Z,U_{1},U_{3}).
\end{align*}
This is because if $Y,Z$ are both uniform and $\mathrm{card}(Y)=\mathrm{card}(Z)$,
we can let $U_{1}$ independent of $(Y,Z)$ and has the same distribution,
and there exists $U_{2},U_{3}$ such that $\mathrm{TRIPLE}(Y,U_{1},U_{2})\,\wedge\,\mathrm{TRIPLE}(X_{1},U_{1},U_{3})$
holds.

Then define the following AEIP that checks whether the random variables
$Y_{1},\ldots,Y_{l},G$ are uniform, and $\prod_{i}\mathrm{card}(Y_{i})=\mathrm{card}(G)$:
\begin{align*}
 & \!\!\!\!\mathrm{PROD}((Y_{1},\ldots,Y_{l});G):\,\\
 & \exists Z^{l},U:\,\bigwedge_{i}(\mathrm{UNIF}_{=}(Y_{i};Z_{i})\,\wedge\,I(Z_{i};Z^{i-1})=0)\\
 & \wedge\,\mathrm{UNIF}_{=}(G;U)\,\wedge\,H(U|Z^{l})=H(Z^{l}|U)=0.
\end{align*}
The following AEIP checks whether $Y,G$ are uniform and $(\mathrm{card}(Y))^{k}=\mathrm{card}(G)$:
\[
\mathrm{POW}_{k}(Y;G):\,\mathrm{PROD}((\underbrace{Y,\cdots,Y}_{k});G).
\]
The following AEIP checks whether $Y,G$ are uniform, $b:=\mathrm{card}(Y)$
is a divisor of $a:=\mathrm{card}(G)$, and $b\ge\sqrt{a}$:
\begin{align*}
 & \!\!\!\!\mathrm{GESQRT}(Y;G):\,\\
 & \exists Z,W,U,V:\,\mathrm{UNIF}_{=}(Y;Z)\\
 & \wedge\,\mathrm{UNIF}(W)\,\wedge\,I(W;Z)=0\\
 & \wedge\,\mathrm{UNIF}_{=}(G;U)\,\wedge\,H(U|Z,W)=H(Z,W|U)=0\\
 & \wedge\,\mathrm{UNIF}_{=}(Z;V)\,\wedge\,H(U|Z,V)=0.
\end{align*}
To check that this AEIP implies $b\ge\sqrt{a}$, note that $\mathrm{card}(V)=b$,
and hence $\mathrm{card}((Y,V))\le b^{2}$. Since $H(U|Y,V)=0$, we
have $a\le b^{2}$. To check that $b\ge\sqrt{a}$ implies this AEIP,
assume $U=(U_{1},U_{2})\sim\mathrm{Unif}([b]\times[a/b])$. Take $Z=U_{1}$,
$V=U_{1}+U_{2}\;\mathrm{mod}\,a$ (assume mod takes values in $[a]$
instead of $[0..a-1]$). We have $Z,V\sim\mathrm{Unif}([b])$ and
$H(U|Z,V)=0$, and the AEIP holds. 

Using this, we can check whether $Y,Z$ are uniform and $\mathrm{card}(Y)\le\mathrm{card}(Z)$:
\begin{align*}
 & \!\!\!\!\mathrm{LE}(Y;Z):\,\\
 & \exists U:\,\mathrm{PROD}((Y,Z);U)\,\wedge\,\mathrm{GESQRT}(Z;U).
\end{align*}
Let $p_{k},q_{k}$ be positive integers such that $\log(k-1)<p_{k}/q_{k}<\log k$
for $k\ge2$. Then $k$ is the only positive integer $i$ satisfying
$2^{p_{k}}\le i^{q_{k}}$ and $i^{q_{k+1}}\le2^{p_{k+1}}$. Hence
we can state $\mathrm{UNIF}_{k}$ as
\begin{align}
 & \!\!\!\!\mathrm{UNIF}_{k}(Y):\,\nonumber \\
 & \exists U,V^{2},W^{2}:\,\mathrm{UNIF}_{2}(U)\,\wedge\,\mathrm{UNIF}(Y)\nonumber \\
 & \wedge\,\mathrm{POW}_{p_{k}}(U;V_{1})\,\wedge\,\mathrm{POW}_{q_{k}}(Y;W_{1})\,\wedge\,\mathrm{LE}(V_{1};W_{1})\nonumber \\
 & \wedge\,\mathrm{POW}_{p_{k+1}}(U;V_{2})\,\wedge\,\mathrm{POW}_{q_{k+1}}(Y;W_{2})\,\wedge\,\mathrm{LE}(W_{2};V_{2}).\label{eq:unif_k}
\end{align}
Therefore, the undecidable AEIP in Theorem \ref{thm:undecidable}
can be stated using only conditional independence constraints and
$\mathrm{UNIF}_{2}$. To complete the proof, note that if $X_{1}\sim\mathrm{Bern}(1/2)$,
then we can check $\mathrm{UNIF}_{2}(Y)$ by $\mathrm{UNIF}_{=}(Y;X_{1})$,
and hence one uniform bit suffices.
\end{IEEEproof}
\medskip{}

As a result, we have the undecidability of the following implication
problem.
\begin{cor}
\label{cor:impl}Fix any integer $r\ge2$. The problem of deciding
whether the following implication holds: $I(X_{A_{j}};X_{B_{j}}|X_{C_{j}})=0$
for all $j\in[m]$ and $\mathrm{card}(X_{1})\le r$ implies $H(X_{1})=0$,
where $A_{j},B_{j},C_{j}\subseteq[n]$, $n$ and $m$ are given, is
undecidable.
\end{cor}
\begin{IEEEproof}
Let $\tilde{p}_{3},\tilde{q}_{3},\tilde{p}_{4},\tilde{q}_{4}$ be
integers such that $\tilde{p}_{3}/\tilde{q}_{3}<\log3<\tilde{p}_{4}/\tilde{q}_{4}$,
and there is no positive integer $k$ such that $\tilde{p}_{3}/\tilde{q}_{3}<(\log k)/(\log3)<\tilde{p}_{4}/\tilde{q}_{4}$.
We define the following AEIP that checks whether $Y$ is uniform and
$\mathrm{card}(Y)\le2$, given that $\mathrm{card}(Y)\le3$:
\begin{align*}
 & \!\!\!\!\mathrm{UNIF}_{\le2|\le3}(Y):\\
 & \exists U,V^{2},W^{2}:\,\mathrm{UNIF}(Y)\,\wedge\,\mathrm{UNIF}(U)\\
 & \wedge\,\mathrm{POW}_{\tilde{p}_{3}}(Y;V_{1})\,\wedge\,\mathrm{POW}_{\tilde{q}_{3}}(U;W_{1})\,\wedge\,\mathrm{LE}(V_{1};W_{1})\\
 & \wedge\,\mathrm{POW}_{\tilde{p}_{4}}(Y;V_{2})\,\wedge\,\mathrm{POW}_{\tilde{q}_{4}}(U;W_{2})\,\wedge\,\mathrm{LE}(W_{2};V_{2}).
\end{align*}
This is similar to \eqref{eq:unif_k} except that $Y,U$ are swapped,
and we have $\mathrm{UNIF}(Y)$ instead of $\mathrm{UNIF}_{2}(Y)$.
Hence, if $\mathrm{card}(Y)=2$, then the AEIP is satisfied by $\mathrm{card}(U)=3$.
If $\mathrm{card}(Y)=3$, then since $\tilde{p}_{3}/\tilde{q}_{3}<(\log k)/(\log3)<\tilde{p}_{4}/\tilde{q}_{4}$
has no solution, the AEIP is not satisfied. 

We show the corollary by a reduction from the decision problem in
Theorem \ref{thm:ci_imp}: 
\begin{align*}
 & \lnot\Big(\exists X^{n}:\,X_{1}\sim\mathrm{Bern}(1/2)\,\wedge\,\forall j\in[m]:I(X_{A_{j}};X_{B_{j}}|X_{C_{j}})=0\Big)\\
\stackrel{(a)}{\Leftrightarrow}\, & \forall X^{n}:\,\Big(\big(X_{1}\sim\mathrm{Bern}(1/2)\,\\
 & \quad\quad\wedge\,\bigwedge_{j\in[m]}(I(X_{A_{j}};X_{B_{j}}|X_{C_{j}})=0)\big)\,\to\,H(X_{1})=0\Big)\\
\stackrel{(b)}{\Leftrightarrow}\, & \forall X^{n},Y:\,\Big(\big(X_{1}\sim\mathrm{Bern}(1/2)\,\wedge\,\mathrm{POW}_{\lfloor\log r\rfloor}(X_{1};Y)\,\\
 & \quad\quad\wedge\,\mathrm{card}(Y)\le r\,\wedge\,\mathrm{UNIF}_{\le2|\le3}(X_{1})\\
 & \quad\quad\wedge\,\bigwedge_{j\in[m]}(I(X_{A_{j}};X_{B_{j}}|X_{C_{j}})=0)\big)\,\to\,H(Y)=0\Big)\\
\stackrel{(c)}{\Leftrightarrow}\, & \forall X^{n},Y:\,\Big(\big(\mathrm{POW}_{\lfloor\log r\rfloor}(X_{1};Y)\,\wedge\,\mathrm{card}(Y)\le r\,\wedge\,\mathrm{UNIF}_{\le2|\le3}(X_{1})\\
 & \quad\quad\wedge\,\bigwedge_{j\in[m]}(I(X_{A_{j}};X_{B_{j}}|X_{C_{j}})=0)\big)\,\to\,H(Y)=0\Big),
\end{align*}
which is a statement in the form given in the corollary, where (a)
is because $P\to Q\Leftrightarrow\lnot P\vee Q\Leftrightarrow\lnot P$
if $P\to\lnot Q$ holds, and (b) is because $\mathrm{card}(Y)\le r$
and $\mathrm{UNIF}_{\le2|\le3}(X_{1})$ are implied by $X_{1}\sim\mathrm{Bern}(1/2)$
and are redundant. To show (c), since $(P\wedge R)\to Q\Leftrightarrow P\to Q$
if $P\to(Q\vee R)$ holds, it suffices to show that $\mathrm{POW}_{\lfloor\log r\rfloor}(X_{1};Y)$,
$\mathrm{card}(Y)\le r$ and $\mathrm{UNIF}_{\le2|\le3}(X_{1})$ implies
$H(Y)=0$ or $\mathrm{card}(X_{1})=2$. If $\mathrm{POW}_{\lfloor\log r\rfloor}(X_{1};Y)$,
$\mathrm{card}(Y)\le r$ and $\mathrm{UNIF}_{\le2|\le3}(X_{1})$,
we have $(\mathrm{card}(X_{1}))^{\lfloor\log r\rfloor}=\mathrm{card}(Y)\le r$,
and $\mathrm{card}(X_{1})\le r^{1/\lfloor\log r\rfloor}$. It is straightforward
to check that $r^{1/\lfloor\log r\rfloor}<4$ if $r\ge2$, and hence
$\mathrm{card}(X_{1})\in\{1,2,3\}$. Since $\mathrm{UNIF}_{\le2|\le3}(X_{1})$
holds, we have $\mathrm{card}(X_{1})\in\{1,2\}$, and the result follows.
\end{IEEEproof}
\medskip{}

Note that Corollary \ref{cor:impl} implies that there does not exist
a program that, when given a list of conditional independence constraints,
outputs the minimum possible $\mathrm{card}(X_{1})$ other than $1$
(i.e., the second smallest possible $\mathrm{card}(X_{1})$; in case
$\mathrm{card}(X_{1})$ can only be $1$, the program should output
$1$). \medskip{}

\subsection{Disjoint Case\label{subsec:disjoint_case}}

We show that that the disjoint case of the conditional independence
implication problem is equivalent to the non-disjoint case.
\begin{thm}
\label{thm:ci_imp_dis}For any $n,m$ and (not necessarily disjoint)
$A_{j},B_{j},C_{j}\subseteq[n]$ for $j\in[0..m]$, there exists $\tilde{n},\tilde{m}$
and disjoint $\tilde{A}_{j},\tilde{B}_{j},\tilde{C}_{j}\subseteq[\tilde{n}]$
for $j\in[0..\tilde{m}]$ such that the implication $\bigwedge_{j\in[m]}(I(X_{A_{j}};X_{B_{j}}|X_{C_{j}})=0)\Rightarrow I(X_{A_{0}};X_{B_{0}}|X_{C_{0}})=0$
holds if and only if the implication $\bigwedge_{j\in[\tilde{m}]}(I(X_{\tilde{A}_{j}};X_{\tilde{B}_{j}}|X_{\tilde{C}_{j}})=0)\Rightarrow I(X_{\tilde{A}_{0}};X_{\tilde{B}_{0}}|X_{\tilde{C}_{0}})=0$
holds. Moreover, the sets $\tilde{A}_{j},\tilde{B}_{j},\tilde{C}_{j}$
can be computed in polynomial time (with respect to the input size
$nm$).
\end{thm}
\begin{IEEEproof}
Two random variables $Y,Z$ are called \emph{perfectly resolvable}
if there exists $X$ such that $I(Y;Z|X)=H(X|Y)=H(X|Z)=0$ \cite{prabhakaran2014assisted}.
Define
\begin{align*}
\mathrm{RESC}(Y,Z,X) & :\,I(Y;Z|X)=H(X|Y)=H(X|Z)=0,
\end{align*}
\[
\mathrm{RES}(Y,Z):\,\exists X:\,\mathrm{RESC}(Y,Z,X).
\]
Note that if $Y,Z$ are perfectly resolvable, then $X$ is uniquely
determined from $Y,Z$ since it is the G{\'a}cs-K{\"o}rner common
part \cite{gacs1973common} of $Y$ and $Z$. 

For each $X_{i}$, let $Y_{i},Z_{i}$ be random variables such that
they are perfectly resolvable and $\mathrm{RESC}(Y_{i},Z_{i},X_{i})$
holds. To prove Theorem \ref{thm:ci_imp_dis}, we will express (not
necesarily disjoint) conditional independence relations $I(X_{A};X_{B}|X_{C})=0$
using disjoint conditional independence relations on $\{Y_{i},Z_{i}\}$. 

First, we redefine $\mathrm{RES}(Y,Z)$ using only disjoint conditional
independence relations. Let
\begin{align*}
\mathrm{RES3}(Y^{3}) & :\,I(Y_{1};Y_{2}|Y_{3})=I(Y_{2};Y_{3}|Y_{1})=I(Y_{3};Y_{1}|Y_{2})=0.
\end{align*}
We will show that $\mathrm{RES}(Y,Z)$ can be equivalently stated
as
\[
\exists U:\,\mathrm{RES3}(Y,Z,U).
\]
The direction $\mathrm{RES}(Y,Z)\Rightarrow\exists U:\,\mathrm{RES3}(Y,Z,U)$
follows from taking $U=X$. To prove the other direction, assume $\mathrm{RES3}(Y,Z,U)$.
Invoke the double Markov property \cite{csiszar2011information}:
\begin{align*}
 & I(U;Y|Z)=I(U;Z|Y)=0\\
\Rightarrow & \exists V:\,H(V|Y)=H(V|Z)=I(Y,Z;U|V)=0.
\end{align*}
This together with $I(Y;Z|U)=0$ implies $I(Y;Z|V)=0$. Hence $\mathrm{RES}(Y,Z)$
follows from taking $X=V$. Note that this also implies $H(V|U)=0$,
and hence $V$ is the G{\'a}cs-K{\"o}rner common part of $X$ and
$Y$, that of $X$and $U$, and also that of $Y$ and $U$.

Write
\[
\mathrm{EQ}(F,G):\,H(F|G)=H(G|F)=0
\]
for the condition that $F,G$ are informationally equivalent. We then
express $\mathrm{EQ}(X_{i},X_{j})$ for distinct $i,j$ using disjoint
conditional independence relations. We will prove that $\mathrm{EQ}(X_{1},X_{2})$
if and only if
\begin{align*}
 & \mathrm{EQRES}(Y_{1},Z_{1},Y_{2},Z_{2}):\\
 & \exists U^{2}:\,\mathrm{RES3}(Y_{1},Z_{1},U_{1}),\,\wedge\,\mathrm{RES3}(Z_{1},U_{1},U_{2})\\
 & \;\;\wedge\,\mathrm{RES3}(U_{1},U_{2},Y_{2})\,\wedge\,\mathrm{RES3}(U_{2},Y_{2},Z_{2}).
\end{align*}
Recall that $\mathrm{RESC}(Y_{i},Z_{i},X_{i})$ holds. For the ``only
if'' part, take $U_{1}=U_{2}=X_{1}$. For the ``if'' part, we have
shown that $\mathrm{RES3}(Y_{1},Z_{1},U_{1})$ implies that the common
part of $Y_{1}$ and $Z_{1}$ is the same as that of $Z_{1}$ and
$U_{1}$, and hence the common part of $Z_{1}$ and $U_{1}$ is also
$X_{1}$. Repeating this argument, we deduce that the common part
of $Y_{2}$ and $Z_{2}$ is also $X_{1}$. The result follows.

We will create 3 copies of $X^{n}$. Let $Y^{3n},Z^{3n}$ satisfy
$\mathrm{EQRES}(Y_{i},Z_{i},Y_{i+n},Z_{i+n})$ for $i\in[2n]$. This
ensures that if $\mathrm{RESC}(Y_{i},Z_{i},X_{i})$ hold for $i\in[3n]$,
then $\mathrm{EQ}(X_{i},X_{i+n})$ for $i\in[2n]$ (i.e., $X_{i},X_{i+n},X_{i+2n}$
are informationally equivalent). After copying, we can express the
(not necesarily disjoint) conditional independence relations $I(X_{A};X_{B}|X_{C})=0$
as $I(X_{A};X_{B+n}|X_{C+2n})=0$, where we write $B+n=\{b+n:\,b\in B\}$,
ensuring the three sets $A,B+n,C+2n$ are disjoint. 

To complete the proof, we express the conditional independence relation
$I(X_{A};X_{B}|X_{C})=0$ ($A,B,C\subseteq[n]$) using disjoint conditional
independence relations on $Y^{3n},Z^{3n}$. Without loss of generality,
we can assume 
\[
I(Y_{i};Y_{[3n]\backslash i},Z_{[3n]\backslash i}|Z_{i})=I(Z_{i};Y_{[3n]\backslash i},Z_{[3n]\backslash i}|Y_{i})=0
\]
for $i\in[3n]$. It is always possible to find such $Y^{3n},Z^{3n}$
given $X^{3n}$ since we can take $Y_{i}=Z_{i}=X_{i}$. Given this
assumption, for $A,B,C\subseteq[n]$, $I(X_{A};X_{B}|X_{C})=0$ holds
if and only if
\[
I(Y_{A};Y_{B+n}|Y_{C+2n})=0.
\]
To show this, first we have $I(X_{A};X_{B}|X_{C})=0$ $\Leftrightarrow$
$I(X_{A};X_{B+n}|X_{C+2n})=0$. Let $a\in A$. Since 
\[
I(Y_{a};Y_{(B+n)\cup(C+2n)},Z_{(B+n)\cup(C+2n)}|Z_{a})=0,
\]
and $H(X_{a}|Z_{a})=I(Y_{a};Z_{a}|X_{a})=0$ (due to $\mathrm{RESC}(Y_{a},Z_{a},X_{a})$),
we have
\[
I(Y_{a};Y_{(B+n)\cup(C+2n)},Z_{(B+n)\cup(C+2n)}|X_{a})=0.
\]
This implies that $I(X_{A};X_{B+n}|X_{C+2n})=0$ $\Leftrightarrow$
$I(X_{A\backslash a},Y_{a};X_{B+n}|X_{C+2n})=0$. Repeating this argument
(note that this argument is also valid for $X_{C+2n}$), we have $I(X_{A};X_{B+n}|X_{C+2n})=0$
$\Leftrightarrow$ $I(Y_{A};Y_{B+n}|Y_{C+2n})=0$. In sum, the implication
$\bigwedge_{j\in[m]}(I(X_{A_{j}};X_{B_{j}}|X_{C_{j}})=0)\Rightarrow I(X_{A_{0}};X_{B_{0}}|X_{C_{0}})=0$
holds if and only if the following implication holds:
\begin{align*}
 & \bigwedge_{i\in[2n]}\mathrm{EQRES}(Y_{i},Z_{i},Y_{i+n},Z_{i+n})\\
 & \wedge\,\bigwedge_{i\in[3n]}\big(I(Y_{i};Y_{[3n]\backslash i},Z_{[3n]\backslash i}|Z_{i})\\
 & \;\;\;\;\;\;\;\;\;\;=I(Z_{i};Y_{[3n]\backslash i},Z_{[3n]\backslash i}|Y_{i})=0\big)\\
 & \wedge\,\bigwedge_{j\in[m]}\big(I(Y_{A_{j}};Y_{B_{j}+n}|Y_{C_{j}+2n})=0\big)\\
 & \Rightarrow\,I(Y_{A_{0}};Y_{B_{0}+n}|Y_{C_{0}+2n})=0.
\end{align*}
Since $\mathrm{EQRES}$ is defined using disjoint conditional independence
relations, the above implication consists of disjoint conditional
independence relations only.
\end{IEEEproof}
\[
\]
We then prove the disjoint case of Corollary \ref{cor:impl} by a
reduction from the non-disjoint case.
\begin{cor}
\label{cor:imp_disjoint}Fix any integer $r\ge2$. The problem of
deciding whether the following implication holds: $I(X_{A_{j}};X_{B_{j}}|X_{C_{j}})=0$
for all $j\in[m]$ and $\mathrm{card}(X_{1})\le r$ implies $I(X_{1};X_{2})=0$,
where the three sets $A_{j},B_{j},C_{j}\subseteq[n]$ are disjoint
for any $j$, and $n\ge2$ and $m$ are given, is undecidable.
\end{cor}
\begin{IEEEproof}
We show a reduction from Corollary \ref{cor:impl}. Let $A_{j},B_{j},C_{j}\subseteq[n]$
be not necessarily disjoint sets. By the construction in Theorem \ref{thm:ci_imp_dis},
letting $Y_{i},Z_{i}$ be random variables such that $\mathrm{RESC}(Y_{i},Z_{i},X_{i})$
holds, we can find disjoint conditional independence relations on
$Y^{n},Z^{n},U^{n'}$ (where $U^{n'}$ are some extra auxiliary random
variables) such that there exists $Y^{n},Z^{n},U^{n'}$ such that
$\mathrm{RESC}(Y_{i},Z_{i},X_{i})$ and these disjoint conditional
independence relations hold if and only if $I(X_{A_{j}};X_{B_{j}}|X_{C_{j}})=0$
for all $j$. The cardinality constraint $\mathrm{card}(X_{1})\le r$
can be enforced by $\mathrm{card}(Y_{1})\le r$ (since $X_{1}$ is
a function of $Y_{1}$). The consequent $H(X_{1})=0$ can be expressed
as $I(Y_{1};Z_{1})=0$. Therefore, the implication ``$\mathrm{RES}(Y_{i},Z_{i})$
for all $i\in[n]$, the aforementioned disjoint conditional independence
relations hold, and $\mathrm{card}(X_{1})\le r$ implies $I(Y_{1};Z_{1})=0$''
holds if and only if the implication ``$I(X_{A_{j}};X_{B_{j}}|X_{C_{j}})=0$
for all $j\in[m]$ and $\mathrm{card}(X_{1})\le r$ implies $H(X_{1})=0$''
holds.
\end{IEEEproof}
\medskip{}

\section{Undecidability of Related Problems\label{sec:related}}

We establish several undecidability results as corollaries of Theorem
\ref{thm:undecidable} and Theorem \ref{thm:ci_imp}.
\begin{cor}
\label{cor:cond_affine}The problem of deciding the truth value of
$\mathbf{v}\in\Gamma_{n}^{*}\,\wedge\,\mathbf{a}^{T}\mathbf{v}\le0\,\wedge\,v_{\{1\}}\le1\,\Rightarrow\,v_{\{1\}}=0$,
where $\mathbf{a}\in\mathbb{Q}^{2^{n}-1}$ and $v_{\{1\}}$ denotes
the entry of $\mathbf{v}$ that represents the entropy of the first
random variable, is undecidable. 
\end{cor}
\begin{IEEEproof}
We show a reduction from the decision problem in Theorem \ref{thm:ci_imp}:
deciding whether there exists random variables $X^{n}$ such that
$I(X_{A_{j}};X_{B_{j}}|X_{C_{j}})=0$ for all $j\in[m]$ and $X_{1}\sim\mathrm{Bern}(1/2)$.
Since $\mathrm{UNIF}(X_{1})$ can be expressed as conditional independence
constraints, assume it is expressed as 
\[
\exists X_{n+1}^{n'}:\,I(X_{A_{m+1}};X_{B_{m+1}}|X_{C_{m+1}})=\cdots=I(X_{A_{m'}};X_{B_{m'}}|X_{C_{m'}})=0.
\]
Since $\mathrm{UNIF}(X_{1})\,\wedge\,0<H(X_{1})\le1$ is equivalent
to $X_{1}\sim\mathrm{Bern}(1/2)$, we have
\begin{align*}
 & \exists X^{n}:\,X_{1}\sim\mathrm{Bern}(1/2)\,\wedge\,\bigwedge_{j\in[m]}(I(X_{A_{j}};X_{B_{j}}|X_{C_{j}})=0)\\
\Leftrightarrow\, & \exists X^{n'}:\,0<H(X_{1})\le1\,\wedge\,\bigwedge_{j\in[m']}(I(X_{A_{j}};X_{B_{j}}|X_{C_{j}})=0)\\
\Leftrightarrow\, & \exists X^{n'}:\,0<H(X_{1})\le1\,\wedge\,\sum_{j=1}^{m'}I(X_{A_{j}};X_{B_{j}}|X_{C_{j}})\le0\\
\Leftrightarrow\, & \lnot\big(\forall X^{n}:\,H(X_{1})\le1\,\wedge\,\sum_{j=1}^{m'}I(X_{A_{j}};X_{B_{j}}|X_{C_{j}})\le0\,\to\,H(X_{1})=0\big),
\end{align*}
which is the negation (the ``$\lnot$'' sign means negation) of
a statement in the form given in Corollary \ref{cor:cond_affine}.
\end{IEEEproof}
\medskip{}

\begin{cor}
\label{cor:cond_affine-1}The problem of deciding the truth value
of $\exists\mathbf{v}\in\Gamma_{n}^{*}:\,\mathbf{a}^{T}\mathbf{v}=0\,\wedge\,v_{\{1\}}=1$,
where $\mathbf{a}\in\mathbb{Q}^{2^{n}-1}$ and $v_{\{1\}}$ denotes
the entry of $\mathbf{v}$ that represents the entropy of the first
random variable, is undecidable.
\end{cor}
\begin{IEEEproof}
We show a reduction from the decision problem in Theorem \ref{thm:ci_imp}.
Using the same notations as Corollary \ref{cor:cond_affine},
\begin{align*}
 & \exists X^{n}:\,X_{1}\sim\mathrm{Bern}(1/2)\,\wedge\,\bigwedge_{j\in[m]}(I(X_{A_{j}};X_{B_{j}}|X_{C_{j}})=0)\\
\Leftrightarrow\, & \exists X^{n'}:\,H(X_{1})=1\,\wedge\,\bigwedge_{j\in[m']}(I(X_{A_{j}};X_{B_{j}}|X_{C_{j}})=0)\\
\Leftrightarrow\, & \exists X^{n'}:\,H(X_{1})=1\,\wedge\,\sum_{j=1}^{m'}I(X_{A_{j}};X_{B_{j}}|X_{C_{j}})=0,
\end{align*}
which is in the form given in Corollary \ref{cor:cond_affine-1}.
\end{IEEEproof}
\medskip{}

\begin{cor}
\label{cor:cond_affine-1-1} The problem of deciding the truth value
of $\forall\mathbf{v}\in\Gamma_{n}^{*}:\,\mathbf{a}_{1}^{T}\mathbf{v}>b_{1}\;\vee\,\cdots\,\vee\;\mathbf{a}_{N}^{T}\mathbf{v}>b_{N}$,
where $\mathbf{a}_{i}\in\mathbb{Q}^{2^{n}-1}$ and $b_{i}\in\mathbb{Q}$,
is undecidable. 
\end{cor}
\begin{IEEEproof}
This can be seen by negating $\exists U^{l}\!:\mathbf{A}\mathbf{h}(U^{l})\preceq\mathbf{b}$.
\end{IEEEproof}
\medskip{}

\section{Acknowledgement}

The author acknowledges support from the Direct Grant for Research,
The Chinese University of Hong Kong (Project ID: 4055133). The author
would like to thank Raymond W. Yeung, Andrei Romashchenko, Alexander
Shen, Milan Studený, Laszlo Csirmaz, Bruno Bauwens and Dariusz Kaloci\'{n}ski
for their invaluable comments.\medskip{}

\medskip{}

\[
\]

\bibliographystyle{IEEEtran}
\bibliography{ref}

\end{document}